\documentclass[aps,prd,showpacs,nofootinbib,floats,floatfix,preprintnumbers,groupedaddress,twocolumn]{revtex4-1}
\usepackage{bm}
\usepackage{latexsym}
\usepackage{dcolumn}
\usepackage{amsmath,amsfonts,amssymb}
\usepackage{graphicx,epsfig}
\usepackage{amsmath}
\usepackage{fancyhdr}
\usepackage{hyperref}
\usepackage{graphicx,epstopdf}
\usepackage{mathrsfs}
\usepackage{url}
\hypersetup{
	colorlinks   = true, %Colours links instead of ugly boxes
	urlcolor     = blue, %Colour for external hyperlinks
	linkcolor    = blue, %Colour of internal links
	citecolor   = red %Colour of citations
}

\begin{document}
\title{Covariant approach to the thermodynamic structure of a generic null surface}
\author{Sumit Dey}
\email{dey18@iitg.ac.in}
\author{Bibhas Ranjan Majhi}
\email{bibhas.majhi@iitg.ac.in}

\affiliation{Department of Physics, Indian Institute of Technology Guwahati, Guwahati 781039, Assam, India}

\date{\today}

%%%%%%%%%%%%%%%%%%%%%%%%%%%%%%%%%%%%%%%%%%%%%%%%
\begin{abstract}
We readdress the thermodynamic structure of geometrical relations on a generic null surface. Among three potential candidates, originated from different components of $R_{ab}$ along the null vectors for the surface (i.e. $R_{ab}q^a_cl^b$, $R_{ab}l^al^b$ and $R_{ab}l^ak^b$ where $q_{ab}$ is the projector on the null surface and  $l^a$, $k^a$ are null normal and corresponding auxiliary vector of it, respectively), the first one leads to Navier-Stokes like equation. Here we devote our investigation on the other two members. We find that $R_{ab}l^al^b$, which yields the evolution equation for expansion parameter corresponding to $l^a$ along itself, can be interpreted as a thermodynamic relation when integrated on the two dimensional transverse subspace of the null hypersurface along with a virtual displacement in the direction of $l^a$.  Moreover for a stationary background the integrated version of it yields the general form of Smarr formula. Although this is more or less known in literature, but a similar argument for the evolution equation of the expansion parameter corresponding to $k^a$ along $l^a$, provided by $R_{ab}l^ak^b$, leads to a more convenient form of thermodynamic relation. In this analysis, contrary to earlier approaches, the identified thermodynamic entities come out to be in covariant forms and also are foliation independent. Hence these can be applied to any coordinate system adapted to the null hypersurface. Moreover, these results are not restricted to any specific parametrisation of $k^a$ and also $k^a$ need not be hypersurface orthogonal. In addition, here any particular dynamical equation for metric is not being explicitly used and therefore we feel that our results are solely based on the properties of the null surface.
\end{abstract}

%%%%%%%%%%%%%%%%%%%%%%%%%%%%%%%%%%%%%%%%

\maketitle

%%%%%%%%%%%%%%%%%%%%%%%%%%%%%%%%%%%%%%%%%%%%%%%%%%%%%%%%%%%%%%%%%%%%%%%%%%%%%%%
\section{Introduction}
The intriguing connection between gravitational dynamics explored on the black hole horizon and classical thermodynamics was laid bare in the seventies following the work of Bekenstein, Hawking and others \cite{Bekenstein:1973ur, Bardeen:1973gs, Hawking:1971vc, Hawking:1976de, Davies:1978mf} (for a review see \cite{Wald:1999vt, Carlip:2014pma, Wall:2018ydq}). This led to the development of the famous black hole mechanics which are a set of intricate equivalences. For every law of black hole mechanics, there exists a corresponding law of classical thermodynamics, thus allowing the black hole to be considered as a thermodynamic object.  However these connections can very well be equivalently established not just only for black hole horizons, but rather on any arbitrary null hypersurface. This allows the attribution of thermodynamical quantities like temperature, entropy etc for any null surface \cite{Parattu:2013gwa}\footnote{Although not fully understood, but there exists certain progress in this direction (see \cite{Chakraborty:2016dwb, Bhattacharya:2018epn, Maitra:2018saa} for few instances).}. It has been shown in literature, that certain projections of the Ricci tensor $R_{ab}$ onto a generic null hypersurface $\mathcal{H}$ and along the corresponding auxiliary null vector assume physical and/or geometrical interpretations. These relevant projections are $R_{ab}l^a l^b$, $R_{ab} l^a k^b$ and $R_{ab} l^a q^{b}_{~c}$ (a related discussion can be followed from Section $3$ of \cite{Chakraborty:2015aja}, see also \cite{Chakraborty:2015hna}). Here $l^a$ represents the null generators of $\mathcal{H}$ and $k^a$ the auxiliary null vector field on $\mathcal{H}$. The induced metric onto $S_t$, where $S_t$ represents the two dimensional spacelike transverse submanifold of $\mathcal{H}$ is given by $q^b_{~c}$ (these notations will be cleared in the next section). Below we first mention the existing discussion of these in order to motivate to our goal.

Damour \cite{Damour:1979wya} (in the context of a black hole event horizon in Einstein gravity) showed that the particular projection, $R_{ab} l^a q^b_{~c}$, on $S_t$ leads to the Damour Navier-Stokes (DNS) equation which is structurally quite similar to the Navier-Stokes (NS) equation. However, the DNS equation can as well be obtained for any generic null hypersurface $\mathcal{H}$ in the spacetime \cite{Gourgoulhon:2005ng, Padmanabhan:2010rp, Kolekar:2011gw} (see Eq. ($6.14$) of  \cite{Gourgoulhon:2005ng} for an excellent review). The DNS equation is an evolution equation that relates the Lie derivative (along $l^a$) of the Hajicek one form with $R_{ab}l^a q^b_{~c}$. The Hajicek one form is a purely geometric quantity defined on $S_t$. 

On the other hand $R_{ab}l^a l^b$ leads to the Null Raychaudhuri (NRC) equation \cite{Poisson:2009pwt}. The NRC equation is a purely geometrical relation which relates the evolution of $\theta_{(l)}$ (the expansion scalar of $l^a$ field) along the null generators $l^a$ with $R_{ab}l^a l^b$. The NRC equation was used as a crucial input by Jacobson to derive the Einstein field equations from the Clausius identity $\delta S = \frac{\delta Q}{T}$, applied on the local Rindler horizons \cite{Jacobson:1995ab}. The Rindler horizons are assumed to be at thermodynamic equilibrium and $\delta Q$ represents the matter energy flux traversing across the causal horizon which results in the change of entropy $\delta S$ (known as Clausius entropy) associated with the horizon. The equilibrium condition requires the crucial restrictive assumption of the vanishing of the second fundamental form and the shear tensor on the null horizon. It is postulated that $\delta S$ is proportional to the area change of the horizon. The above formalism was extended  to the non equilibrium case, in the regime of which, the shear tensor and the expansion scalar on the null surface cannot be set to zero \cite{Eling:2006aw, Chirco:2009dc, Dey:2017fld}.
Gravitational equations for certain modified theories of gravity were also obtained from such similar thermodynamic considerations \cite{Chirco:2009dc}. Later this concept of Clausius entropy was extended to arbitrary bifurcate null surface \cite{Baccetti:2013ica} and the Einstein equations were also derived for stretched light cone \cite{Parikh:2017aas}. Moreover, Jacobson \cite{Jacobson:2015hqa} derived the Einstein field equations as applied to local causal diamonds (constructed at any point in the spacetime) by extremizing the total entanglement entropy of the null horizon and the matter inside of it.

Now we concentrate on $R_{ab}l^a k^b$. It was shown by Padmanabhan and his collaborators \cite{Padmanabhan:2002sha, Kothawala:2007em, Chakraborty:2015aja} that a certain projection of the Einstein equation (specifically $G_{ab}l^a k^b$ and hence $R_{ab}l^a k^b$) yields a thermodynamic interpretation which is structurally similar to the first law of thermodynamics. The main difference between Padmanabhan's \cite{Padmanabhan:2002sha, Kothawala:2007em, Chakraborty:2015aja} and Jacobson's \cite{Jacobson:1995ab} approaches in order to relate thermodynamics is the choice of the component of the Einstein equation. For Jacobson, the relevant projection component is $R_{ab}l^al^b$, whereas for Padmanabhan, the choice is $R_{ab}l^a k^b$. In fact, it is pointed out in \cite{Chakraborty:2015aja} that the neater component to consider is $R_{ab}l^a k^b$ which produces the thermodynamic identity without any restrictive assumptions like the vanishing of the second fundamental form and the shear tensor on $\mathcal{H}$ (which was a crucial assumption in Jacobson's approach). The argument behind this is $R_{ab}l^ak^b$ picks out the component of $R_{ab}l^a$ along $l_a$, the null generators which are intrinsic to the null surface $\mathcal{H}$. Whereas the other one corresponds to that of $R_{ab}l^a$ along $k^a$ (see section $3$ of \cite{Chakraborty:2015aja} for more details). Padmanabhan's approach has been generalized to the case of Lanczos-Lovelock theories of gravity \cite{Paranjape:2006ca, Kothawala:2009kc, Chakraborty:2015wma} as well.

%Out of the three projections, only $R_{ab} l^a q^b _{~c}$ admits a physical interpretation (an evolution equations for the Hajicek one form i.e DNS equation) that is explicitly \textcolor{blue}{inherent to the null structure} of $\mathcal{H}$ without taking recourse to the specific gravitational equation of motion. 

However the existing physical interpretations of $R_{ab}l^a l^b$ and $R_{ab}l^a k^b$, so far, have been made explicit via the specific gravitational equation of motion. 
%That is, for $R_{ab}l^a l^b$, the thermodynamic interpretation has been brought about in the (restrictive) equilibrium case as applied to local Rindler horizons for Einstein gravity. For $R_{ab}l^a k^b$, the thermodynamic interpretation has been brought about without any such restrictions of equilibrium condition. 
Moreover, in the latter case the same has been done for a generic null hypersurface by invoking the adapted Gaussian null coordinates (GNC) \cite{Friedrich:1998wq, morales2008second, Parattu:2015gga, Racz:2007pv}. This makes the identified expression of the thermodynamics entities to be in ``noncovariant'' form. In this paper, we aim to investigate whether in a completely covariant fashion $R_{ab}l^a l^b$ and $R_{ab}l^a k^b$ can be provided any physical interpretation, without invoking any specific gravitational dynamics; i.e. solely based on the properties of null surface $\mathcal{H}$. In our analysis, the NRC equation (for both the $l^a$ and $k^a$ vector fields) is the starting point in providing a physical interpretation for the two concerned projections. We show that the underlying dynamics of the background is not ``explicitly'' necessary to provide such interpretation.

%Since the NRC is an evolution equation and is a purely geometric relationship \textcolor{blue}{on $\mathcal{H}$}, we are able to provide a physical interpretation to the projections $R_{ab}l^a l^b$ and $R_{ab}l^a k^b$ without the specification of the gravitational equation of motion. That is, our generic null hypersurface may evolve under any theory of gravity, not just the Einstein Hilbert or the Lanczos-Lovelock theory (\textcolor{blue}{or any modified theories of gravity}). Whatever be the specific theory of gravity, once we are provided an arbitrary null surface in the spacetime, we in our analysis are able to provide a physical interpretation to both the projections $R_{ab}l^a l^b$ and $R_{ab}l^a k^b$.

Our analysis and results are divided in different sections. In section \ref{Foliation} we provide a brief overview of the null foliation of the spacetime manifold in the neighbourhood of our generic null hypersurface $\mathcal{H}$, which acts as the building block of the present analysis. In section \ref{NRCl}, we analyse if $R_{ab}l^a l^b$ can be provided a physical interpretation. We begin with the NRC equation for $l^a$  and then integrate it on the transverse spacelike surface $S_t$.  Performance of  a virtual displacement along the null generators of $\mathcal{H}$ leads to a possibility of thermodynamic identity. We feel that this is not a surprising result at all as NRC equation is being used in search of thermodynamics of horizon and therefore it possesses such an inherent structure. Still we present this one in order to provide a segue into our main topic of providing a physical interpretation to $R_{ab}l^a k^b$ in a covariant way. In going through the steps we shall observe few interesting features of the approach which are probably not emphasized in literature.  
It is noticed that in the special case of a stationary black hole system, the expression of the energy is related to the well known Komar energy $E_K$ (please see  Chapter $4$, page number $149$ of \cite{Poisson:2009pwt}). Moreover the integrated form of the thermodynamic identity leads to a generalised form of Smarr formula \cite{Smarr:1972kt} which, as given in literature, is of the form $E_K=2ST$, Here  $T$ is the temperature of the horizon (see \cite{Padmanabhan:2003pk, Padmanabhan:2009kr, Banerjee:2010yd, Banerjee:2010ye} for discussions related to this identity).

 Next we concentrate our attention to $R_{ab}l^ak^b$ in Section \ref{rablakb}. Invoking the NRC equation corresponding to $k^a$ (the auxiliary null vector field) and integrating it on the transverse space $S_t$ and allowing for the virtual displacement along $k^a$, we arrive at a thermodynamic interpretation of $R_{ab}l^a k^b$ which is structurally equivalent to the first law of thermodynamics.  The null foliation (though non-unique) of the spacetime manifold allows us to have a completely covariant expression of the energy term. This is because the expression of the energy term contains geometrical quantities defined on the null surface $\mathcal{H}$. These geometrical quantities once defined on $\mathcal{H}$ will be independent of the foliation chosen.  Here we provide our definition of the {\it ``geometrical work function"} in order to make way for the thermodynamic identity independent of any theory of gravity.
 
  Previously equivalent thermodynamic interpretation (analogous to the first law of thermodynamics) have been provided in the Einstein-Hilbert \cite{Chakraborty:2015aja} and Lanczos-Lovelock theory \cite{Chakraborty:2015wma}. However there are certain important differences between the work in \cite{Chakraborty:2015aja},\cite{Chakraborty:2015wma} and ours. In \cite{Chakraborty:2015aja} and \cite{Chakraborty:2015wma}, the derivations of the thermodynamic identity have been performed near a generic null hypersurface without any assumed symmetries of the spacetime. However the derivations have been performed with respect to (w.r.t) an adapted null coordinate system constructed in the neighbourhood of the generic null surface $\mathcal{H}$ known as the GNC system. One noticeable feature of such a construction is that the expression of the energy is compatible to the GNC metric only. For the GNC construction, the auxiliary null vector field $k^a$ is affinely parametrized and hypersurface orthogonal. We in our case however foliate the spacetime in the vicinity of the generic null surface $\mathcal{H}$ by a family or stack of null hypersurfaces. Then allowing for the $3+1$ induced foliation of the family of the null surfaces, we derive exactly the same structural thermodynamic identity in a completely coordinate independent fashion. The reason as to why we are able to achieve this covariantly is mentioned in section \ref{Foliation} once we introduce the construction of the null foliated spacetime. Our construction does not require $k^a$ to be affinely parametrized and hypersurface orthogonal. Another difference in our approach from that adopted in earlier ones is that our starting point is the NRC equation for $k^a$, whereas no such equation has been explicitly used in these works.  It may be pointed out that the work function (or pressure) in \cite{Chakraborty:2015aja, Chakraborty:2015wma, Kothawala:2010bf} has constantly been defined as $P = -T_{ab}l^a k^b$ i.e owing entirely from the matter energy tensor. The entropy density has then been defined as the Bekenstein-Hawking entropy density for the Einstein-Hilbert case \cite{Chakraborty:2015aja} and as the the Wald entropy density for the Lanczos-Lovelock models \cite{Chakraborty:2015wma}. However in our interpretation, we have identified what we call as the ``{\it geometrical work function" (or geometrical pressure)}, entirely from geometrical quantites. In analogy to the entanglement entropy, we call our identified entropy density as the ``entanglement entropy density'' since it depends on the geometry of the relevant surface. Under the umbrella of such an interpretation, we have aimed to provide the thermodynamic identification independent of any theory of gravity.
 
For the reader, we summarize the structural and interpretational difference between the approach in \cite{Chakraborty:2015aja, Chakraborty:2015wma} and ours.
\begin{itemize}
\item The thermodynamic identity in \cite{Chakraborty:2015aja, Chakraborty:2015wma} and is brought through the GNC construction while ours is brought about through a $3+1$ foliation of the null family $\mathcal{H}_\Phi$.
%\item The GNC construction requires $k^a$ to be both affinely parametrized as well as hypersurface orthogonal. We do not impose any such restrictions.
\item  The expression of the energy in the GNC is solely adapted to these coordinates. On the contrary ours is in a covariant form and hence can be applied to any structure of the null surface.
\item In \cite{Chakraborty:2015aja, Chakraborty:2015wma}, the pressure or work function has been consistently defined w.r.t the matter energy tensor $P = -T_{ab}l^a k^b$, while we define a so called ``geometrical work function".
\item The entropy density in \cite{Chakraborty:2015aja} is the Bekenstein Hawking entropy density, while it is the Wald entropy density for the Lanczos-Lovelock models \cite{Chakraborty:2015wma}. For our interpretation, the entropy density is consistently the entanglement entropy density irrespective of the theory of gravity.
\end{itemize}

Finally we shall conclude in Section \ref{Conclusion}. We use the appendices to show that the covariant expression of the energy we derive entirely matches with the expression of the energy obtained via the GNC system in \cite{Chakraborty:2015aja} for the Einstein-Hilbert case. There we also provide outlines of derivations for expressions used in the text.

%Having looked at all these relevant projection components consider the following points. 
%%\item $R_{ab}l^a q^b_{~c}$ : \quad This results in the DNS evolution equation which relates the Lie derivatve (along $l^a$) of the Hajicek one form $\Omega_a$ to $R_{ab}l^a q^b_{~c}$.  Without invoking any specific gravitational field equations we obtain an evolution equation for the Hajicek one form which is a purely geometrical quantity defined on $S_t$.
%\end{itemize}
%%%%%%%%%%%%%%%%%%%%%%%%%%%%%%%%%%%%%%%%%%%%%%%%%%%%%%%%%%%%%%%%%%%%%%%%%%%%%%%%%%%%%%
Before proceeding ahead, we state that throughout the paper we have the metric signature $(-,+,+,+)$. We use units where $\hbar$ and $c$ are set to unity. The lowercase Latin indices $a$,$b$,.. represent the bulk spacetime indices and run from $0$ to $3$. The Greek symbol $\mu$ run from $1$ to $3$ and represents the spatial coordinates on a spacelike hypersurface. The transverse coordinates in the spacelike subspace $S_t$ orthogonal to $l^a$ and $k^a$ are represented by the upper case Latin indices $A$,$B$,.. and run from $2$ to $3$.
%%%%%%%%%%%%%%%%%%%%%%%%%%%%%%%%%%%%%%%%%%%%%%%%%%%%%%%%%%%%%%%

\section{Brief overview of our geometrical construction and comparison with  GNC}\label{Foliation}
In this section we briefly mention the geometrical setup of the arena where we aim to find the thermodynamical connections. All our analysis will be focused on a generic null hypersurface $\mathcal{H}$ in the spacetime manifold $(\mathcal{M},g)$ and we aim to establish thermodynamical laws w.r.t this null surface. For details of the construction, we refer the reader to \cite{Gourgoulhon:2005ng} and hence we provide a brief overview here. 

Our generic null hypersurface is submanifold in the spacetime such that the induced metric $q_{ab}$ of $\mathcal{H}$ is degenerate. The null surface can be defined by the constant value of a scalar field $\Phi$ such that the normal one-form to the hypersurface $\mathcal{H}$ is defined as,
\begin{equation}\label{oneforml}
l_a = e^{\rho} \nabla_a \Phi ~,
\end{equation}
with $\rho$ being a scalar function on null surface. This entails the Frobenius identity and hence is hypersurface orthogonal to $\mathcal{H}$. The null generators $l^a$ of this hypersurface satisfy the geodesic equation,
\begin{equation}
l^i \nabla_i l^a = \kappa l^a ~,
\label{nonaffinity}
\end{equation}
where $\kappa$ is the non-affinity parameter and therefore these are not affinely parametrized. 

Now we add structure to the null hypersurface $\mathcal{H}$ by the auxiliary null foliation in its neighbourhood. The aim of this construction is geared to the fact that we want $l_a$ to be defined not just only on $\mathcal{H}$, but rather in its open neighbourhood (in $(\mathcal{M},g)$). This allows us to define covariant derivative of  $l_a$ in the entire spacetime, which would not have been the case had we been restricted only on $\mathcal{H}$. The construction proceeds by foliating the spacetime in the neighbourhood of $\mathcal{H}$, by a family of null hypersurfaces parametrized by different (constant) values of the scalar field $\Phi$. Thus the family of the null hypersurfaces foliating the spacetime is denoted by $\mathcal{H}_{\Phi}$. Our generic null hypersurface $\mathcal{H}$ is then nothing, but a particular member of this set, say $\mathcal{H}_{\Phi = 1}$. Then (\ref{oneforml}) stays valid even in this construction of a family of null hypersurfaces. Therefore ``extends" the domain of scalar field $\rho$ from the null surface $\mathcal{H}$ to its open neighbourhood where the foliation takes place. One important caveat of such a construction is that even though  such a foliation $\mathcal{H}_{\Phi}$ is non-unique, all geometrical quantities are independant of the foliation. That is to say, once a geometrical quantity has been evaluated at $\mathcal{H}$, its remains the same no matter what construction is chosen for $\mathcal{H}_{\Phi}$ to foliate the open neighbourhood of $\mathcal{H}$.

In order to define the projector onto the null surface we require an auxiliary null vector $k^a$ satisfying $l^a k_a = -1$ and $k^a k_a =0$. However the auxiliary null field $k^a$ satisfying the above two conditions is not uniquely defined. We then proceed to foliate the null hypersurface $\mathcal{H}$ (or the family $\mathcal{H}_{\Phi}$) by spacelike slices in the spirit of $3+1$ splitting. We consider a family of $t = \text{constant}$ spacelike hypersurfaces $\Sigma_{t}$ (parametrized by $t \in \Bbb R$) that intersect the family $\mathcal{H}_{\Phi}$ on $2$ dimensional spacelike hypersurfaces $S_{t,\Phi}$; i.e. 
$S_{t,\Phi} := \mathcal{H}_{\Phi} \cap \Sigma_t$.
Then for our concerned hypersurface $\mathcal{H}$, $S_t$ surfaces are precisely
$S_t = S_{t,\Phi = 1} = \mathcal{H} \cap \Sigma_t$. Now out of the family of the non-unique null auxiliary field $k^a$, we choose the one that is orthogonal to the surface $S_t$ and hence satisfies the conditions $k^ak_a =0$, $k^a l_a = -1$
 and $k_a e^a_{~A} = 0$. Here $\{e^a_{~A}\}$ refers to the basis of the tangent space $\mathcal{T}_p (S_t)$ established on $(S_t,q)$. Hence in our construction we have the vector fields $l^a$ and $k^a$ orthogonal to $\mathcal{T}_p (S_t)$ i.e $l^a q_{ab} = 0$ and $k^a q_{ab} = 0$.
 We now mention the structural differences between the above construction and the GNC adapted (to $\mathcal{H}$) system (a detailed discussion on the construction of GNC can be followed from \cite{Friedrich:1998wq, morales2008second, Parattu:2015gga, Racz:2007pv}). This discussion is important in order to identify the underlying constructional difference in thermodynamic interpretation from the earlier attempts which rely on the structure of GNC.  In this construction with the coordinates $(u,r,x^A)$, the generic null hypersurface is stationed at $r=0$. The null normal to the hypersurface $\mathcal{H}$ is defined as the gradient of the $r = \text{constant}$ surfaces i.e $l_a = \partial_a r$ and are non-affinely parametrized satisfying (\ref{nonaffinity}). The auxiliary null vector field $k^a$ in the GNC is by construction chosen to be along affinely parametrized null geodesics. That is, we move away from the null surface stationed at $r=0$ along the ingoing null geodesic of $k^a$. In the GNC, the auxiliary null vector field $k^a = -(\partial/\partial r)^a$ has the affine parameter $r$ and points along the direction of decreasing $r$ (ingoing). It can also be seen that the null geodesic $k_a$ is hypersurface orthogonal to the $u=\text{constant}$ surfaces i.e $k_a = -\partial_a u$. Hence we see that the coordinates adapted to the null surface $\mathcal{H}$ at $r=0$ are $(u,x^A)$. As a result of this adapted coordinatization $(u,r,x^A)$ of $(\mathcal{M},g)$ in the vicinity of $\mathcal{H}$, the thermodynamic interpretation of $R_{ab}l^a k^b$ comes explicitly as GNC dependant. However in the construction of the null foliated spacetime introduced above provided with a $3+1$ foliation, we do not impose any coordinatization. That is, we just demand that the neighbourhood of $\mathcal{H}_{\Phi = 1}$ is null foliated by a family of hypersurfaces $\mathcal{H}_{\Phi = c}$. This null foliated spacetime is then foliated by $t=\text{constant}$ spacelike surfaces. That is all what we require for providing $R_{ab}l^a l^b$ and $R_{ab}l^a k^b$ a thermodynamic interpretation. This is because as we have mentioned earlier, even though the foliation is non-unique, yet all geometrical quantities relevant to $\mathcal{H}$ are independant of the foliation chosen. In fact, if we want, we can ``adapt" a coordinate system w.r.t to $\mathcal{H}$. A very famous example would be a coordinate system that  is ``stationary w.r.t the null hypersurface $\mathcal{H}$" \cite{Gourgoulhon:2005ng}. For example associated with the spacelike foliation of $(\mathcal{M},g)$ via $t=\text{constant}$ surfaces, we can consider a $3+1$ coordinate system $(x^a = (t,x^{\mu}))$. Here $t$ is the coordinate associated with the time development vector $t^a = (\frac{\partial}{\partial t})^a = (1,0,0,0)$ and $x^{\mu}$ are the spacelike coordinates on the $t = \text{constant}$ slice. $t^a$ can be expressed in terms of the lapse function $N$ and the shift vector $b^a$ as,
\begin{equation}
t^a = Nn^a + b^a ~,
\end{equation} 
where $n^a$ is the timelike unit normal to $\Sigma_t$. The time development vector basically connects  neighbouring slices of $\Sigma_t$  and $\Sigma_{t+ dt}$ with the same spatial coordinates.  
If in this coordinate system $(x^a)$, the equation of $\mathcal{H}$ does not depend on the coordinate $t$ and only involves the spatial coordinates $x^{\mu}$, then $(x^a)$ is a coordinate system that is stationary w.r.t $\mathcal{H}$. This implies that there exists a scalar function $\psi(x^1,x^2,x^3)$ such that say $\psi(x^1, x^2, x^3) =1$ defines $\mathcal{H}$. This means that the location of the $2$-surface $S_t$ is fixed on the $t = \text{constant}$ surface ( $\Sigma_t$ with the coordinates $x^{\mu} = (x^1,x^2,x^3)$). It can then be shown \cite{Gourgoulhon:2005ng} that for such a stationary coordinate system $x^a = (t, x^{\mu})$ adapted to $\mathcal{H}$, we have,
 \begin{equation}
 l^a \overset{\mathcal{H}}{=} t^a + V^a ~,
 \label{stationary111}
\end{equation} 
where (\ref{stationary111}) is valid only on the null hypersurface. $V^a$ then represents the surface velocity of $\mathcal{H}$ w.r.t this adapted stationary coordinates. Had we proceeded with finding the physical interpretations of $R_{ab}l^a l^b$ and $R_{ab}l^a k^b$ w.r.t to such an adpated coordinate system, then our expressions would turn out to be coordinate dependant. Just like in GNC construction, the generic null surface $r=0$ is intersected by the $u = \text{constant}$ null surfaces. However these scalar functions are themselves adapted to define the coordinate system in the neighbourhood of $\mathcal{H}$. Adapting any coordinate system w.r.t $\mathcal{H}$ would defeat our purpose of providing the interpretations covariantly. We will show in our computations that nowhere do we require the information of the lapse function $N$ and the shift vector $b^a$ associated with the $3+1$ foliation. Nor do we as advertised, require the need of a coordinate system adapted to $\mathcal{H}$. In fact, anticipating one step further, whatever physical interpretation we covariantly attest to $R_{ab}l^a l^b$ and $R_{ab}l^a k^b$ should in fact be independent of our null foliated construction itself (since it is non-unique). The interpretation ought to be specific only to the null hypersurface $\mathcal{H}$.

 In our construction, the auxiliary null vector field need not be along null geodesics as well as hypersurface orthogonal. The auxiliary null vector field in our case is by construction an ingoing normal to the spacelike $2$-surfaces $S_t$ and hence extends out into the open neighbourhood of $\mathcal{H}$. It can be shown (see \cite{Gourgoulhon:2005ng} for details), that $k_a$ satisfies the following relation,
\begin{equation}\label{kspace}
\begin{split}
\partial_a k_b - \partial_b k_a = \frac{1}{2N^2} \Big[\partial_a \Big(\ln(\frac{N}{M})\Big) l_b - \partial_b \Big(\ln(\frac{N}{M})\Big)l_a\Big] ~,
\end{split}
\end{equation}
where $M = \frac{e^{\rho}}{N}$. The relation (\ref{kspace}) shows that $k^a$ is not hypersurface orthogonal (as on the right hand side (RHS) $l_a$ appears). Hence $k^a$ is not the generator of any null hypersurface or in other words, the hyperplane normal to $k^a$ is not integrable into a smooth submanifold surface. It can also be shown (see \cite{Gourgoulhon:2005ng}) that,
\begin{equation}\label{kspace1}
\begin{split}
k^i \nabla_i k_a = -\frac{1}{2 N^2} \Pi^i_{~a} \partial_i \ln\Big(\frac{N}{M}\Big) ~,
\end{split}
\end{equation}
where $\Pi^i_{~a} = \delta^i_{~a} + k^i l_a$ is the projection tensor onto $\mathcal{H}$ along $k_i$. The relation (\ref{kspace1}) essentially shows that the auxiliary null vector field $k^a$ does not satisfy the geodesic equation.

After mentioning the constructional difference we are now aiming to our main goal. This will require several relevant quantities, which we shall list below. The detailed geometric interpretation of them are provided in \cite{Gourgoulhon:2005ng}.
The projection tensor onto $S_t$ is,
\begin{eqnarray}
q_{ab} = g_{ab} + l_a k_b + k_a l_b ~.
\label{inducedmetrix}
\end{eqnarray}
The second fundamental form $\Theta_{ab}$ of the null surface $\mathcal{H}$ and its irreducible decomposition in terms the expansion scalar $\theta_{(l)}$ and the traceless shear tensor $\sigma_{ab}$ corresponding to $l^a$ are,
\begin{equation}
\Theta_{ab} =\frac{1}{2}q^i_{~a} q^k_{~b}\pounds_{l} q_{ik} = (\nabla_i l_k )q^i_{~a} q^k_{~b} 
= \frac{1}{2} q_{ab} \theta_{(l)} + \sigma_{ab} ~.
\end{equation}
The rotation one form $\omega_a$ satisfies the following relation,
\begin{equation}
\omega_a l_b = \nabla_a l_b - \Theta_{ab} + l_a (k^i \nabla_i l_b)~.
\end{equation}
The Hajicek $1$ form $\Omega_a$ is basically the projection of the rotation one-form onto $S_t$ and is given by,
\begin{eqnarray}
\Omega_a = q^i_{~a} \omega_i 
= \omega_a + \kappa k_a ~.
\end{eqnarray}
The transversal deformation rate tensor $\Xi_{ab}$ of $S_t$ projected on to the tangent plane of ($S_t$,q) along the auxiliary null vector field $k^a$ and its irreducible decomposition into the traceless part $(\theta_{(k)})$ and the shear tensor $(\sigma_{(k)_{ab}})$ are given by,
\begin{equation}
\Xi_{ab} = \frac{1}{2} q^i_{~a} q^j_{~b} \pounds_k q_{ij} = q^i_{~a} q^j_{~b} \nabla_i k_j
= \frac{1}{2} q_{ab} \theta_{(k)} + \sigma_{(k)_{ab}}~. 
\end{equation}
The covariant derivative defined on $(S_t,q)$ for any spatial vector $v_a$ confined to it is,
\begin{eqnarray}
{^2 \mathcal{D}_a} v_b = q^i_{~a} q^k_{~b} \nabla_i v_k ~.
\end{eqnarray}
The Ricci tensor corresponding to the $2$-surface spacelike manifold $(S_t,q)$ is designated by $^{2}R_{ab}$ and its corresponding Ricci scalar as $^{2}R = q^{ab}~^{2}R_{ab}$. 
%%%%%%%%%%%%%%%%%%%%%%%%%%%%%%%%%%%%%%%%%%%%%%
\section{$R_{ab}l^al^b$: a thermodynamic identity ?}\label{NRCl}
The quantity $R_{ab}l^al^b$ is best represented by NRC equation for the null vector $l^a$.
Usually, NRC equation is being used to explore thermodynamical behaviour of black hole horizon. Both for proving the area increase theorem as well as finding first law of black hole mechanics, this plays a central role. It must be noted that in all these analysis NRC equation came in the middle of the calculation and always applied for Killing vector which is null on the horizon. For instance, see the discussion around Eq. $(8.168)$ to Eq. $(8.173)$ of \cite{padmanabhan2010gravitation}. Moreover, in Jacobson's analysis \cite{Jacobson:1995ab} this has been used to derive the Einstein's equation of motion from the Clausius relation. Therefore, apparently NRC equation has an inherent thermodynamical structure on the horizon. Although it is not explicitly mentioned in literature, but the way it has been used one can immediately identify this property. In all these earlier analysis, the expressions for thermodynamical entities are taken as input at the very beginning and then finally NRC equation is used to obtain required conclusion. Also, as we mentioned earlier, this is strictly confined to Killing case (or asymptotically Killing).  

Here we shall take the ``opposite'' route. We will begin with the NRC equation for an arbitrary null vector $l^a$ (say), not necessarily a Killing one. Interestingly, the integration of this on the two dimensional subspace on which $l^a$ is normal, leads to first law of thermodynamics like structure. This analysis has some noticeable features. First of all, this is valid for any arbitrary null surface and so the results are valid beyond Killing vector field. Secondly, a more general expression of gravitational energy can be obtained. Finally, this thermodynamic structure is the property of the null surface, instead of being related to any associated gravitational theory. 
% \textcolor{red}{Finally, obtention of first law does not need the Einstein's equations of motion. In this sense we call this as ``off-shell'' analysis.} 

Let us now start our calculation. The NRC equation for $l^a$ is given by \cite{Poisson:2009pwt, Gourgoulhon:2005ng}
\begin{equation}
\frac{d\theta_{(l)}}{d\lambda_{(l)}} = \kappa\theta_{(l)}-\frac{1}{2}\theta_{(l)}^2 - \sigma_{ab}\sigma^{ab}  - R_{ab}l^a l^b~,
\label{B1}
\end{equation}
where $l^a$, satisfying (\ref{nonaffinity}), is parametrised by non-affine parameter $\lambda_{(l)}$, i.e. $l^a = dx^a/d\lambda_{(l)}$. We now make a virtual displacement of the null hypersurface along its own generators by an amount $\delta \lambda_{(l)}$.
Multiplying both sides of equation (\ref{B1}) by the transverse elementary area $dA=\sqrt{q}d^2x$ of $S_t$ and $\delta\lambda_{(l)}$ and then integrating $dA$ only one finds
\begin{eqnarray}
\delta\lambda_{(l)}\int_{S_t}dA~\kappa\theta_{(l)} &=& \delta\lambda_{(l)}\int_{S_t}dA~\Big[\frac{d\theta_{(l)}}{d\lambda_{(l)}}
\nonumber
\\
&+&\frac{\theta_{(l)}^2}{2}+\sigma_{ab}\sigma^{ab}+R_{ab}l^al^b\Big]~.
\label{B2}
\end{eqnarray}
Now since we know that 
\begin{equation}
\theta_{(l)}= \frac{1}{\sqrt{q}}\frac{d\sqrt{q}}{d\lambda_{(l)}}~,
\label{B3}
\end{equation}
the term on the left hand side (LHS) of (\ref{B2}) can be expressed as follows:
\begin{eqnarray}
\frac{1}{8 \pi G} \delta\lambda_{(l)}\int_{S_t}dA~\kappa\theta_{(l)} &=& =  \int_{S_t} d^2x \frac{\kappa}{2 \pi} \delta \lambda_{(l)} \frac{d}{d \lambda_{(l)}}\Big(\frac{1}{4 G}\sqrt{q}\Big) \nonumber \\
&=& \int_{S_t} d^2 x T \delta_{\lambda(l)} s ~. 
\label{B4}
\end{eqnarray}
In the above we introduced a factor $1/(8\pi G)$. Here we have identified $T = \frac{\kappa}{2 \pi}$ as the temperature of the null surface and is hence related to the non affinity parameter $\kappa$ of the null generators of the null hypersurface $\mathcal{H}$. We here also identify $s = \frac{\sqrt{q}}{4G}$ as the entropy density of the null surface. This, in analogy to entanglement entropy, we may interpret as  entanglement entropy density (more will be discussed on this analogy in the next section). 
In the same way, multiplying the RHS of (\ref{B2}) by the numerical factor $(1/8\pi G)$, the resultant equation can be interpreted as the following thermodynamic identity:
\begin{equation}
\int_{S_t }d^2x T\delta_{\lambda(l)}s = \delta_{\lambda(l)}E ~,
\label{B5}
\end{equation}
where $E$ is the energy associated to the null surface, given by
\begin{equation}
E=\frac{1}{8\pi G}\int d\lambda_{(l)}\int_{S_t}dA~\Big[\frac{d\theta_{(l)}}{d\lambda_{(l)}}
+\frac{\theta_{(l)}^2}{2}+\sigma_{ab}\sigma^{ab}+R_{ab}l^al^b\Big]~.
\label{B6}
\end{equation}
Note that in the whole analysis, we never used Einstein's equations of motion and so the result is very generic to any null surface. The virtual displacement $\delta \lambda_{(l)}$ is consistent with any physical process that virtually displaces the $2$-surface $S_t$ along the null generators itself say from positions $\lambda_{(l)} = 0$ to $\lambda_{(l)} = \delta \lambda_{(l)}$.

 We can however provide an alternative interpretation to the NRC equation (\ref{B1}) under the process of the virtual displacement $\delta \lambda_{(l)}$. We first multiply both sides of Eq. (\ref{B1}) by the transverse area element of the $2$-surface $S_t$ together with a multiplicative factor of $\frac{1}{8 \pi G}$ i.e $\frac{1}{8\pi G} dA = \frac{1}{8 \pi G}\sqrt{q}d^2 x$ and the virtual displacement $\delta \lambda_{(l)}$. We then integrate the resulting equation over $S_t$,
 \begin{eqnarray}
 &\frac{1}{8 \pi G}\delta\lambda_{(l)}\int_{S_t}d^2 x \sqrt{q}~\kappa\theta_{(l)} \nonumber \\ 
 &=\frac{1}{8 \pi G} \delta \lambda_{(l)} \int_{S_t} d^2 x \sqrt{q} ~\Big(-\frac{\theta_{(l)}^2}{2} + \sigma_{ab} \sigma^{ab} \Big) \nonumber \\   
 &+ \frac{1}{8 \pi G}\delta \lambda_{(l)} \int_{S_t} d^2 x \sqrt{q} \Big[\frac{d \theta_{(l)}}{d \lambda_{(l)}} + \theta^2_{(l)} + R_{ab}l^a l^b\Big] ~.
 \label{nrcl1}
 \end{eqnarray}
 Following Eq. (\ref{B4}), the LHS can be identified as, 
 \begin{eqnarray}
 \frac{1}{8 \pi G}\delta\lambda_{(l)}\int_{S_t}d^2 x \sqrt{q}~\kappa\theta_{(l)} = \int_{S_t} d^2 x T \delta_{\lambda(l)} s ~,
 \label{entangle}
 \end{eqnarray}
 where $s = \frac{\sqrt{q}}{4 G}$ is the entropy density of the null hypersurface and will be identified as the entanglement entropy density.
  Looking at the first term of the RHS of (\ref{nrcl1}), we find the integrand contains the well known dissipation term $\mathcal{D}$ corresponding to the null hypersurface,
  \begin{eqnarray}
  \mathcal{D} = \Theta_{ab}\Theta^{ab} - \theta_{(l)}^2 = -\frac{1}{2} \theta^2_{(l)} + \sigma_{ab}\sigma^{ab} ~.
  \label{dissipation}
  \end{eqnarray}
  The identification of the dissipation term basically comes from the $R_{ab}l^a q^b_{~c}$ component, which results in the DNS equation. The physical interpretation of this viscous dissipation term and its observer dependence has been explored in \cite{Padmanabhan:2010rp}. Finally, if we identify the variation of the energy under $\delta \lambda_{(l)}$ virtual displacement as,
  \begin{eqnarray}
  \delta_{\lambda (l)} E = \frac{1}{8 \pi G}\delta \lambda_{(l)} \int_{S_t} d^2 x \sqrt{q} \Big[\frac{d \theta_{(l)}}{d \lambda_{(l)}} + \theta^2_{(l)} + R_{ab}l^a l^b\Big] ~,
  \label{varenergy}
    \end{eqnarray}
we can then identify (\ref{nrcl1}) as ,
\begin{eqnarray}
\int_{S_t }d^2x T\delta_{\lambda(l)}s =   \frac{1}{8 \pi G} \delta \lambda_{(l)} \int_{S_t} d^2 x \sqrt{q} \mathcal{D} + \delta_{\lambda(l)}E ~.
\label{thermidenalt}
\end{eqnarray}    
The first term on the RHS of  (\ref{thermidenalt}) is identified as the heat generation part under the virtual displacement $\delta \lambda_{(l)}$ process due to irreversible viscous dissipation effects of the null surface $\mathcal{H}$, 
\begin{eqnarray}
	\delta_{\lambda(l)} Q_{\text{dis}} = \frac{1}{8 \pi G} \delta \lambda_{(l)} \int_{S_t} d^2 x \sqrt{q} \mathcal{D}~. 
	\label{irrev}
\end{eqnarray} 
A very interesting feature of this dissipation term $\mathcal{D}$ is that it comprises entirely of geometrical quantities. Hence it can be thought that the dissipative sector of the heat generation basically arises due to gravitational energy fluxes through the null surface. A similar interpretation was previously put forth in \cite{Chirco:2009dc}. The authors of \cite{Chirco:2009dc} show via an analogy between the horizon null congruence and a continuous medium that the dissipative heat generation is purely due to gravitational effects. Our expression of the internal heat production (\ref{irrev}) for non-affine generators $l^a$ matches with Eq. ($47$) of \cite{Chirco:2009dc} for non-affine Killing vector field approximately generating the Killing horizon, provided we set $\theta_{(l)}$ to zero in our case. The authors in \cite{Chirco:2009dc} constructed  a local Rindler wedge in the neighbourhood of a spacetime point $P$. The Rindler horizon is approximately generated by the Killing vector field $\xi^a$ which satisfies $\xi^a \nabla_a \xi^b = \kappa \xi^b$. The Killing vector field is tangent to the affinely parametrized null congruence $l^a$ of the Rindler horizon. The viscous dissipative part of the heat generation term or \textit{uncompensated heat} in Eq ($47$) of \cite{Chirco:2009dc} contains the norm $||\hat{\sigma}||$, where $\hat{\sigma}_{ab}$ is the shear tensor corresponding to the Killing congruence. However in our construction, the arbitrary null surface is generated by the non-affinely paramaterized $l^a$ field and hence all geometric quantities in our expression (\ref{irrev}) of the  dissipative heat generation pertains to the $l^a$ congruence itself.  %\textcolor{red}{They further show that the irreversible entropy generation coincides with the famous Hawking-Hartle tidal heating term (for Einstein gravity??). They further propund that this internal entropy generation is related to the work done by the perturbative tidal field on the horizon. We borrow the spirit of \cite{}, to have (\ref{irrev}) interpreted as the work done by the viscous dissipative forces on the null surface $\mathcal{H}$. However such a work has its origin in the internal/microscopic degrees of freedom. Hence (\ref{irrev}) cannot be thought of the reversible work done by the viscous forces on the null surface.}
Under such an identification, we have from Eq. ($\ref{thermidenalt}$),
\begin{eqnarray}
	\int_{S_t }d^2x T\delta_{\lambda(l)}s  =  \delta_{\lambda(l)} Q_{\text{dis}} + \delta_{\lambda{(l)}}E ~.
	\label{thermidenalt2}
\end{eqnarray}
 
We now carry over this analysis to the special case of a {\it stationary} black hole system, for example the Kerr spacetime. In that case, the non affinity parameter $\kappa$ is independent of the transverse coordinates $\{x^A\}$ and hence can be taken outside the integral in equation (\ref{B4}). This allows us to interpret equation (\ref{B5}) in the more familiar form of the thermodynamic identity (first law of thermodynamics),
\begin{equation}\label{stat}
T \delta_{\lambda(l)} S = \delta_{\lambda(l)} E ~,
\end{equation}
where $S = \frac{A}{4 G}$ is the entropy of the null hypersurface and hence proportional to its area $A$.
Let us now investigate the expression (\ref{B6}) of the  energy $E$ for the stationary black hole case. In this case $\mathcal{H}$ is the event horizon of the black hole and $l^a$ is the timelike Killing vector which is null on $\mathcal{H}$. Therefore, we denote $l^a$ as $l^a\equiv\xi^a=(1,0,0,\Omega_H)$ where $\Omega_H$ is the angular velocity of the black hole. Then, the first term on the RHS of (\ref{B6}) can be evaluated as follows. Integration over $\lambda_{(l)}$ yields $\int_{S_t} dA~\theta_{(l)}\Big|_{\lambda_{(l)}=1}^{\lambda_{(l)}=2}$ and since the value of $\theta_{(l)}$ vanishes at the stationary point on horizon $S_t$, this term will not contribute. In this case, for the ``quasi static physical process'' the next two terms of RHS can be neglected on $S_t$ compared to the other terms (known as equilibrium or near-equilibrium situation \cite{Jacobson:1995uq, Chatterjee:2011wj}). Therefore the energy expression (\ref{B6}), for this special case reduces to
\begin{equation}
E= \frac{1}{8\pi G}\int d\lambda_{(l)}\int_{S_t}dA~R_{ab}\xi^a\xi^b~.
\label{B7}
\end{equation}
Also analogously we investigate the second alternative interpretation provided via Eq. (\ref{thermidenalt2}), through the introduction of the viscous dissipative part of the heat generation under the ``quasi-static physical process'' for the stationary black hole system. This process actually displaces the $2$-surface $S_t$ initially at $\lambda_{(l)} = 0$ (say) to the position $\lambda_{(l)} = \delta \lambda_{(l)}$. However for the stationary black hole case, the virtual displacement is through a quasi-equilibrium process. That is, initially the $2$-surface $S_t$ is at equilibrium at $\lambda_{(l)} = 0$ and then via a quasi-static process it is displaced to the stationary equilibrium state at $\lambda_{(l)} = \delta \lambda_{(l)}$. As a result of this quasi-static virtual displacement process the dissipative heat generation $\delta_{\lambda(l)} Q_{\text{dis}}$ is basically zero. Under such a process for the stationary black hole, we hence have (\ref{stat}). The energy expression for the stationary black hole case in this alternative interpretation still remains the same and is given by (\ref{B7}).

This expression of the energy (\ref{B7}) known to be proportional to the Komar expression for conserved quantity (see \cite{Poisson:2009pwt} for details on Komar conserved quantity), calculated on the horizon. The volume element on $\mathcal{H}$ is $d\Sigma^a=-\xi^a d\lambda_{(l)}dA$. Also, we can express $R_{ab}\xi^b$ as $R_{ab}\xi^b = [\nabla_b,\nabla_a]\xi^b = (1/2)\nabla^b(\nabla_a\xi_b-\nabla_b\xi_a)\equiv (1/2)\nabla^b J_{ab}$, where in the last step Killing equation $\nabla_a\xi_b+\nabla_b\xi_a=0$ has been used. Note that $J_{ab}$ can be identified as the Noether potential for Einstein gravity. Using all these in (\ref{B7}) one obtains
\begin{eqnarray}
E= - \frac{1}{16\pi G}\int_{\mathcal{H}} d\Sigma_a \nabla_b J^{ab}=-\frac{1}{32\pi G} \oint_{S_t} d\Sigma_{ab} J^{ab}~.
\label{B8}
\end{eqnarray}
In the last equality, the Stoke's law for integration has been applied.
Now the Komar conserved quantity is defined as \cite{Poisson:2009pwt}
\begin{equation}
E_{K} = - \frac{1}{16\pi G} \oint_{S_t} d\Sigma^{ab}J_{ab}~,
\label{B9}
\end{equation}
which for Killing vector corresponding to time transnational symmetry, gives mass term $M_H$ at the horizon  while that for Killing vector corresponding to azimuthal symmetry leads to $-2\Omega_H J_H$ where $J_H$ is the angular momentum at $S_t$. Therefore (\ref{B9}) for stationary black hole, like Kerr metric, yields $E_K=M_H-2\Omega_H J_H$. Comparison of (\ref{B8}) and (\ref{B9}) yields $E=(1/2)E_K$.

Next we can integrate Eq. (\ref{stat}) over $\lambda_{(l)}$. Since for stationary background $l^a=\xi^a$ and $\xi^a\nabla_a \kappa=0$, we have
\begin{equation}
\delta_{\lambda{(l)}} T= \delta \lambda_{(l)}\frac{d T}{d\lambda_{(l)}} = \delta\lambda_{(l)}\xi^a\nabla_a T=0~.
\label{B10}
\end{equation}
Therefore integration of (\ref{stat}) yields $E=TS$ and since $E=(1/2)E_K$, in terms of Komar conserved quantity on horizon, one finds
\begin{equation}
E_K=2TS~.
\label{B11}
\end{equation}
This has already been shown in literature \cite{Padmanabhan:2003pk, Padmanabhan:2009kr, Banerjee:2010yd, Banerjee:2010ye} that it is the general form of famous Smarr formula \cite{Smarr:1972kt} (for a particular dynamical black hole, similar relations have also been achieved \cite{Majhi:2014hpa}). For instance, in the case of Kerr, the above leads to our well known Smarr expression $M_H = 2\Omega_HJ_H+(\kappa A/4\pi G)$. This shows that the integrated form of NRC equation on the stationary horizon along the Killing vector is the Smarr relation. 
%%%%%%%%%%%%%%%%%%%%%%%%%%%%%%%%%%%%%%%%%%

\section{$R_{ab}l^ak^b$, favourable candidate for thermodynamic interpretation: a covariant approach}\label{rablakb}
We are now in a position to hit the better candidate among the different projections of $R_{ab}$; i.e. $R_{ab}l^ak^b$ (the logic for better choice has been discussed in \cite{Chakraborty:2015aja}) which serves as a thermodynamic identity. The approach will be similar to the earlier section and hence the outcome will be covariant in nature.
We start with the following evolution equation of the transversal deformation rate tensor $\Xi_{ab}$ along the null generators $l^a$ of the hypersurface $\mathcal{H}$:
\begin{eqnarray}
	q_a^i q_b^j \pounds_l \Xi_{ij} &=& \frac{1}{2}(^{2}\mathcal{D}_a \Omega_b + ^2\mathcal{D}_b \Omega_a) + \Omega_a \Omega_b -\frac{1}{2} {^{2}R_{ab}} 
	\nonumber
	\\
	&+& \frac{1}{2} q_a^i q_b^j R_{ij} -\Big(\kappa + \frac{\theta_{(l)}}{2}\Big) \Xi_{ab} 
	\nonumber
	\\
	&-& \frac{\theta_{(k)}}{2} \Theta_{ab} + \Theta_{ai} \Xi^i_b + \Xi_{ai}\Theta^i_b ~.
	\label{transverse_evol}
\end{eqnarray}
The detailed derivation of this is given in \cite{Gourgoulhon:2005ng}. Taking trace on both sides we obtain the following identity:
\begin{eqnarray}
-\kappa \theta_{(k)} &=  \Big(- {^2\mathcal{D}_a }\Omega^a - \Omega_a \Omega^a +  \theta_{(l)} \theta_{(k)} +l^i \nabla_i \theta_{(k)} + \frac{1}{2}{^2 R} \Big) 
\nonumber
\\
 &-    \Big(R_{ij}l^i k^j+\frac{1}{2} R\Big) ~.
\label{thermoevolution_trace}
\end{eqnarray}
See Appendix \ref{App1} for the derivation of this. 

Similar to NRC for $l^a$, equation (\ref{thermoevolution_trace}) can be interpreted as the evaluation equation for $\theta_{(k)}$ along $l^a$. Another point to be noted is that one can write $l^a\nabla_a \theta_{(k)}=k^a\nabla_a\theta_{(l)}$ and then this provides the directional derivative of $\theta_{(l)}$ along $k^a$. This has been sketched in Appendix \ref{App2}.
Below we shall show taking inspiration from  the earlier section, the NRC equation (\ref{thermoevolution_trace}) via the virtual displacement along $k^a$ can also be provided an interpretation as a thermodynamic relation on the $2$-surface, without explicitly invoking the underlying dynamical equation for gravity. 

The auxiliary null vector field $k^i$ is parametrised as $k^i = - (dx^i/d \lambda_{(k)})$, where $\lambda_{(k)}$ is the parameter along the $k^i$ field. Note that negative sign is chosen here in the parametrisation. The reason is as follows. Usually the null vector $l^a$ is chosen to be outgoing and so $x^a$ increases along this direction. Whereas the auxiliary vector $k^a$ is regarded as ingoing one and hence $x^a$ decreases along this field. Now here we are interested in the thermodynamic interpretation of (\ref{thermoevolution_trace}) when evaluated along $k^a$. In this case, to identify the relevant thermodynamic entities like entropy, energy, etc. in its usual meaning it is required to define change of $x^a$ along $k^a$ as positive one. Therefore we have the coordinate variation under the virtual displacement $\delta \lambda_{(k)}$ as $\delta x^a=-k^a \delta\lambda_{(k)}$. The physical interpretation of this displacement has been explained in \cite{Chakraborty:2015aja} and we briefly mention it here. Let us consider two null surfaces under the null based foliation of the spacetime by the family $\mathcal{H}_{\Phi}$. The null surfaces have to be the solutions to the specific theory of gravity that we are considering {\it implicitly}. Let us suppose that the null surfaces are stationed at $\lambda_{(k)} = 0$ and $\lambda_{(k)} = \delta \lambda_{(k)}$. The virtual displacement $\delta \lambda_{(k)}$ is essentially a process that lets us shift from one solution of the hypersurface to the other since $k^a$ is an ingoing vector.
Then the expansion of the congruence of the auxiliary null vector field in terms of rate of area element of $S_t$ is given as,
\begin{eqnarray}\label{d15}
\theta_{(k)} & =& q^{ij}(\nabla_i k_j) = \frac{1}{2} q^{ij}\pounds_{k} q_{ij} = \frac{1}{\sqrt{q}}\pounds_k \sqrt{q}
\nonumber
\\
& =& -\frac{1}{\sqrt{q}}\frac{d}{d \lambda_{(k)}} \sqrt{q} ~.
\end{eqnarray}
The details of this relation has been sketched in Appendix \ref{totalderq}.
Now we multiply both sides of (\ref{thermoevolution_trace}) with $\delta \lambda_{(k)}$ and the elemental area $\sqrt{q} d^2 x$ on the $2$-surface $S_t$ along with an overall factor of $\frac{1}{8 \pi G}$. Then the integration over the $2$-surface yields,
\begin{equation}\label{d16}
\begin{split}
&-\delta \lambda_{(k)} \int_{S_t} d^2 x \sqrt{q} \frac{\kappa}{2 \pi} \frac{1}{4G}\theta_{(k)} \\
&= \delta \lambda_{(k)} \int_{S_t} d^2 x \sqrt{q} \frac{1}{8 \pi G}\Big[\frac{1}{2}{^2 R} + k^i \nabla_i \theta_{(l)} -\Omega_a \Omega^a - {^2 \mathcal{D}}_A \Omega^A\Big]\\
& -\delta \lambda_{(k)} \int_{S_t} d^2 x \sqrt{q}\frac{1}{8 \pi G}\Big[R_{ij}l^i k^j+\frac{1}{2} R \Big]~.
\end{split}
\end{equation}
The LHS of above equation (\ref{d16}) can be rewritten in the following form:
\begin{equation}\label{d17}
\begin{split}
&-\delta \lambda_{(k)} \int_{S_t} d^2 x \sqrt{q} \frac{\kappa}{2 \pi} \frac{1}{4G}\theta_{(k)}\\
&= \int_{S_t} d^2 x \frac{\kappa}{2 \pi} \delta \lambda_{(k)} \frac{d}{d \lambda_{(k)}}\Big(\frac{1}{4 G} \sqrt{q}\Big) = \int_{S_t} d^2 x T \delta_{\lambda(k)} s ~,
 \end{split}
\end{equation}
where we associate the temperature $T$ of the null surface $S_t$ as being related to the non-affinity parameter via $T =(\kappa/2 \pi)$. The entropy density $s$ of the null surface is identified to be $s =(\sqrt{q}/4G)$. We identify this entropy density defined on the null hypersurface to be the entanglement entropy density. We will have more to say on the nature of this entropy density shortly. 

Now focusing on the first term on the RHS of (\ref{d16}), we identify it as the variation of the energy associated with the null surface $S_t$ under the virtual displacement   $\delta \lambda_{(k)}$, i.e
\begin{equation}\label{d18}
\begin{split}
\delta_{\lambda(k)} E =& \delta \lambda_{(k)} \int_{S_t} d^2 x \sqrt{q} \frac{1}{8 \pi G}\Big[\frac{1}{2}{~^2 R} \\
& + k^i \nabla_i \theta_{(l)} + \theta_{(l)}\theta_{(k)} - \Omega_a \Omega^a - {~^2\mathcal{D}}_A \Omega^A\Big]~.
\end{split}
\end{equation}
Performing an indefinite integration over $\lambda_{(k)}$ allows us to have an expression for the energy associated with the $2$-surface $S_t$,
\begin{equation}\label{energy111}
\begin{split}
E &= \int d\lambda_{(k)} \int_{S_t} d^2 x \sqrt{q} \frac{1}{8 \pi G}\Big[\frac{1}{2}{~^2 R} \\
& + k^i \nabla_i \theta_{(l)} + \theta_{(l)}\theta_{(k)} - \Omega_a \Omega^a - {~^2\mathcal{D}}_A \Omega^A\Big]~.
\end{split}
\end{equation}
Before proceeding ahead, we note that the expression of the energy as obtained in (\ref{energy111}) is reminiscent of the Hawking-Hayward energy definition \cite{Hayward:1993ph,  Prain:2015tda}. Our aim here is to show that the analogous Null Raychaudhuri equation for the auxiliary null vector field $k^i$ (\ref{thermoevolution_trace}) has a thermodynamic interpretation under the process of the virtual displacement $\delta \lambda_{(k)}$. That is, we proceed towards interpreting (\ref{d16}) as a thermodynamic identity. To this end we have identified the LHS of (\ref{d16}) as $T \delta _{\lambda(k)} s$ integrated on the null $2$-surface $S_t$ and the first term on the RHS of (\ref{d16}) as the variation of the energy of the null surface under the virtual displacement. The thermodynamic identity would be complete if we are allowed to interpret the second term of of (\ref{d16}) as the virtual work done  under the displacement of the null surface $S_t$ by $\delta \lambda_{(k)}$. Allowing ourselves the liberty, we identify the ``geometric work function'' associated with the virtual displacement $\delta\lambda_{(k)}$ as $P = -1/(8 \pi G)(R_{ij}l^i k^j+\frac{1}{2} R)$. 
%The motivation behind such a description will be made evident in the next section where we consider a specific coordinate system adapted to the null surface, in which $(-T_{ij}l^i k^j)$ will turn out to have the interpretation of the transverse pressure associated with the virtual work done under the displacement $\delta\lambda_{(k)}$.%
Following this, we have,
\begin{eqnarray}\label{d20}
&&-\delta \lambda_{(k)} \int_{S_t} d^2 x \sqrt{q}\frac{1}{8 \pi G}\Big(R_{ij}l^i k^j+\frac{1}{2} R\Big)
\nonumber
\\ 
&&= \delta \lambda_{(k)} \int_{S_t} d^2 x \sqrt{q} P
\equiv F \delta \lambda_{(k)} ~,
\end{eqnarray}
where $F$ is the integral of the work function over the transverse space $S_t$ of the null surface and  $F \delta \lambda_{(k)}$ is to be interpreted as the virtual work done under $\delta \lambda_{(k)}$. Combining (\ref{d17}), (\ref{d18}) and (\ref{d20}), we see that (\ref{d16}) can be succinctly formulated as,
\begin{equation}\label{d21}
\int_{S_t} d^2 x T \delta_{\lambda(k)} s = \delta_{\lambda(k)} E + F \delta \lambda_{(k)} ~.
\end{equation}
We remind that this interpretation holds for all the variations that are consistent with the virtual displacement. That is to physically interpret this, let us say that our virtual displacement is a ``physical" process that virtually shifts our null surface $\mathcal{H}$ from say $\lambda_{(k)} = 0$ to $\lambda_{(k)} = \delta \lambda_{(k)}$. Under such a virtual variation process energy flows through the null surface $\mathcal{H}$. The energy is given by $\delta_{\lambda(k)} E$. The energy then contributes to the heat energy $\int_{S_t} d^2 x T \delta_{\lambda(k)} s$ and the virtual work done $F \delta \lambda_{(k)}$ under this virtual displacement process.

We further note that the expression of the energy (\ref{energy111}) can be rewritten as, 
\begin{equation}\label{d19}
\begin{split}
E &= \frac{1}{2}\int d\lambda_{(k)} \Big(\frac{\chi}{2 G}\Big)\\
&+\frac{1}{8 \pi G}\int d\lambda_{(k)} \int_{S_t} d^2 x \sqrt{q} \Big[ k^i \nabla_i \theta_{(l)} + \theta_{(l)}\theta_{(k)}\\
& - \Omega_a \Omega^a - {^2\mathcal{D}}_A \Omega^A\Big] ~,
\end{split}
\end{equation} having noted that $\chi$ represents the following integral over $S_t$ (a $2$ dimensional manifold) defined as,
\begin{equation}\label{Eulerch}
\chi  = \frac{1}{4 \pi} \int_{S_t} d^2 x \sqrt{q} {~^2R}~.
\end{equation}
If the transverse space $S_t$ of the null surface is compact then, $\chi$ is precisely equal to the Euler characteristic of the $S_t$; if not, then $\chi$ is defined via the integral (\ref{Eulerch}). For example if the topology of $\mathcal{H}$ is $\mathbb{R} \times  \mathbb{S}^2$, then $S_t$ is the compact surface $\mathbb{S}^2$ and hence $\chi$ then represents the Euler characteristics of the sphere.

Now,  we restrict to the special case where the non-affinity parameter $\kappa$ and hence the temperature $T$ associated with the null surface $S_{t}$ is independent of the transverse coordinates of $S_{t}$ (for example a stationary black hole system). In that sense $T$ can be taken outside the integral, and then identifying the total change of the entropy $S$ of the null surface under the virtual displacement  as $\delta_{\lambda(k)} S$ =  $\int_{S_t} d^2 x  \delta_{\lambda(k)} s$ , we have further simplification of (\ref{d21}),
\begin{equation}\label{d22}
T\delta_{\lambda(k)} S = \delta_{\lambda(k)} E + F \delta \lambda_{(k)} ~.
\end{equation}

Before proceeding ahead towards showing the equivalence of our approach with previous results and interpretations let us mention the differences instead first. We iterate that our analysis is independent of any gravitational theory as per se. For example we have not invoked the Einstein's field equations or any other gravitational field equation for that matter in our interpretation of the Null Raychaudhuri equation for $k^i$ as providing a thermodynamic identity. This is in contrast to previous results, which have been specifically formulated for explicit theories of gravity \cite{Chakraborty:2015aja,Kothawala:2010bf, Chakraborty:2015wma}. This is precisely the reason as to why we define a gravitational/geometric  work function $P = -1/(8 \pi G)(R_{ij}l^i k^j+\frac{1}{2} R)$ as opposed to $P = -T_{ab}l^a k^b$ as is done in previous works \cite{Chakraborty:2015aja,Kothawala:2010bf, Chakraborty:2015wma} which identify the work function or pressure entirely in terms of the matter energy momentum tensor. This also entails as to why irrespective of any gravitational theory, we have identified the entropy density of the null hypersurface under the virtual displacement process to be the entanglement entropy density. The observer under such a virtual displacement process is the null observer moving along the integral curves of the null generators $l^a$. We assume that our generic null hypersurface $\mathcal{H}$ actually partitions the spactime into timelike and spacelike regions. Then the quantum fields living in spatial slices on both these two sides can be entangled. The degrees of freedom (dof) of the quantum fields in the spacelike acausal region is not accessible to a timelike observer in the timelike causal region. The timelike observer then calculates the reduced density matrix by tracing out the dof of the quantum fields on the acausal side. The entanglement entropy is then defined as the Von Nuemann entropy of this reduced density matrix. By introducing a momentum cut-off the entanglement entropy is shown to be proportional to the area of the null surface $\mathcal{H}$. Since the entropy density introduced in our case is proportional to $\sqrt{q}$, with an analogy to entanglement entropy, we propose that this as entanglement entropy density as measured by the null observer, moving along $l^a$. In this regard we mention that a similar concept has been taken by  Jacobson \cite{Jacobson:1995ab,Eling:2006aw,Jacobson:2015hqa} at the very beginning of his analysis in order to obtain the Einstein's equation by extremizing the entropy of the Rindler horizon as well as for a causal diamond.

Let us mention again, that in our analysis leading towards the thermodynamic interpretation (under the virtual displacement $\delta \lambda_{(k)}$), (\ref{d21}) is independent of any coordinate system as opposed to \cite{Chakraborty:2015aja, Chakraborty:2015wma}, which produces the equivalent thermodynamic identity, but under the null adapted GNC coordinates. A specific requirement under the GNC system is the fact that there is only one null hypersurface stationed at the position $r=0$. This null hypersurface partitions the spacetime between timelike and spacelike regions. However in our case, we have foliated the spacetime in the neighbourhood of $\mathcal{H}$ by a family of null hypersurfaces $\mathcal{H}_\Phi$ and have focussed on producing the theroralesmodynamic identity on any one of them, say $\mathcal{H}_{\Phi=1} = \mathcal{H}$. A specific advantage of such a foliation is that all the relevant geometrical quantities that can be defined on the null surface (for example expansion scalar, second fundamental form etc) are independent of the foliation. Another requirement specific to the GNC analysis is that the auxiliary null vector field is affinely parametrized i.e $k_a = - \partial_a u$ or in other words $k^a$ is hypersurface orthogonal to $u=\text{constant}$ surfaces. This, we believe is a certain restriction on the analysis. We can however do away with such a restriction. Under the system of the foliation of spacetime introduced in \ref{Foliation}, we do not require $k^a$ to be affinely parametrized and hypersurface orthogonal. In fact, under this general structure, the hyperplane normal to $k^a$ cannot be integrated into some integrable surface. As a result of such a null foliation of the spacetime in the vicinity of $\mathcal{H}$, our interpretation of the energy and the work function pertain entirely to geometric quantities defined in the spacetime manifold. The way we have made a distinction between the energy term and the work function is to identify that the energy expression (\ref{energy111}) contains terms that are defined on the null $2$-surface $S_t$ along with a term involving the directional derivative of such quantities defined on $S_t$. However the work function (\ref{d20}) contains terms that are defined for the entire spacetime manifold. In doing so, we have obtained a covariant (but foliation based) expression of the energy of the null surface. Previous expositions \cite{Chakraborty:2015aja, Chakraborty:2015wma} into the energy term under the purview of providing a thermodynamic interpretation have however come under the context of an adapted coordinate system w.r.t a fiduciary null surface i.e the GNC construction. As a result, previous such descriptions of the energy have been coordinate dependant. 

As a mathematical curiosity, the above relation (\ref{thermoevolution_trace}) can also be derived in an alternative method following \cite{Kothawala:2010bf}. In \cite{Kothawala:2010bf}, the thermodynamic identity analogous to equation (\ref{d22}) was shown for static null horizons. We generalize the results to any arbitrary null hypersurface via the use of two following relations. The first one relates the Ricci tensor ($^{2}R_{ab}$) of the $2$-dimensional transverse Riemannian manifold $(S_t,q)$ with the $4$-dimensional Riemann curvature tensor of the spactime $(\mathcal{M},g)$, the second fundamental form $\Theta_{ab}$ of the null hypersurface $\mathcal{H}$ and the transversal deformation rate $\Xi_{ab}$ (see \cite{Gourgoulhon:2005ng}),
\begin{eqnarray}
^{2}R_{ab} &= q^{~i}_{a} q^{~j}_{b} q^{~k}_{c} R^{c}_{~jki} - \Xi_{ab} \theta_{(l)} - \Theta_{ab} \theta_{(k)} \nonumber \\
 &+ \Theta_{a}^{~c} \Xi_{cb} + \Xi_{a}^{~c} \Theta_{cb} ~.
 \label{eqn15}
\end{eqnarray}
Taking the trace of the above equation, we obtain,
\begin{eqnarray}
^{2}R = q^{ij}q^{ck} R_{cjki} - 2 \theta_{(l)}\theta_{(k)} + 2 \Theta^{ab}\Xi_{ab} ~.
\label{eqn16}
\end{eqnarray}
Using the definition of the orthogonal projection tensor onto the $2$-surface $S_t$ as  $q^{ab} = g^{ab} + l^a k^b + k^a l^b$, we have,
\begin{eqnarray}
q^{ij}q^{ck} R_{cjki} = R + 4 R_{ij} l^i k^j + 2 R_{abcd}l^a k^b k^c l^d ~.
\label{eqn17}
\end{eqnarray}
Similarly using the irreducible decomposition of both the second fundamental form of $\mathcal{H}$ and the traversal deformation tensor of the two surface $S_t$ i.e $\Theta_{ab} = \frac{1}{2} \theta_{(l)}q_{ab} + \sigma_{ab}$ and $\Xi_{ab} = \frac{1}{2} \theta_{(k)} q_{ab} + \sigma_{(k)ab}$, we get,
\begin{eqnarray}
2 \Theta_{ab} \Xi^{ab} = \theta_{(l)} \theta_{(k)} + 2 \sigma_{ab} \sigma_{(k)}^{ab} ~.
\label{eqn18}
\end{eqnarray}
Upon using the equations (\ref{eqn17}) and (\ref{eqn18}) in equation (\ref{eqn16}), we obtain as a result,
\begin{eqnarray}
R &= {^{2}R} - 4 R_{ij}l^i k^j - 2 R_{abcd}l^a k^b k^c l^d \nonumber \\
&+ \theta_{(l)}\theta_{(k)} - 2\sigma_{ab}\sigma_{(k)}^{ab} ~.
\label{eqn19}
\end{eqnarray}
The second relation that will be put to use is,
\begin{eqnarray}
R_{abcd}l^a k^b k^c l^d &= {^{2}\mathcal{D}_a \Omega^a} + \Omega_a \Omega^a - \kappa \theta_{(k)} - l^i \nabla_i \theta_{(k)} \nonumber \\
& -\frac{1}{2}\theta_{(l)} \theta_{(k)} - \sigma_{ab}\sigma_{(k)}^{ab} - R_{ab}l^a k^b ~.
\label{eqn20}
\end{eqnarray}
A detailed derivation of the above result is provided in appendix \ref{proof}. In fact the relation (\ref{eqn20}) can be regarded as a generalization to  Eq. ($5$) of \cite{Kothawala:2010bf} (which is valid only for static null horizons) to any arbitrary null surface. Now simply the use of equation (\ref{eqn20}) in equation (\ref{eqn19}) leads us to equation (\ref{thermoevolution_trace}). 

Let us conclude this section by mentioning that in Appendix (\ref{existing_results}), we show the equivalence of the energy term (\ref{energy111}) and the gravitational/geometric work function term with those obtained in \cite{Chakraborty:2015aja} via the GNC construction under the purview of Einstein gravity.
%%%%%%%%%%%%%%%%%%%%%%%%%%%%%%%

\section{\label{Conclusion}Conclusion} In the present work, we have investigated as to whether the components $R_{ab}l^a l^b$ and $R_{ab}l^a k^b$ can be provided any physical interpretation in a covariant fashion. Our starting point in both the cases have been the NRC equation (for $l^a$ and $k^a$ fields), which is a covariantly formulated geometrical relationship involving the evolution of a particular geometrical quantity with either $R_{ab}l^a l^b$ or $R_{ab}l^a k^b$.

For $R_{ab}l^a l^b$, we started with the NRC for the null generators $l^a$ (\ref{B1}), and then provided a virtual displacement $\delta \lambda_{(l)}$. We then integrated the resulting equation onto the transverse spacelike $2$-surface $S_t$ and obtained  relevant thermodynamical structures (\ref{B5}) and (\ref{thermidenalt2}). We have provided two alternative interpretations of the resulting thermodynamic identity. In the first interpretation (\ref{B5}), we identified that under the virtual displacement process $\delta \lambda_{(l)}$, an amount of energy $\delta_{\lambda(l)} E$ sweeps across through the null surface $\mathcal{H}$. The expression of the energy is provided in (\ref{B6}). This energy flow results in the heat exchanged as a result of the entropy variation of the null surface. The temperature of $\mathcal{H}$ is associated with the non-affinity parameter $\kappa$ of the null generators and the entropy density is proportional to $\sqrt{q}$ of the area element of $S_t$. In our second interpretation (\ref{thermidenalt2}), we have identified the energy variation $\delta_{\lambda(l)}E$ and the irresversible heat $\delta_{\lambda(l)} Q_{\text{dis}}$ that flows (as a result of the virtual displacement process) across through the null surface $\mathcal{H}$. These quantities are given by ($\ref{varenergy}$) and ($\ref{irrev}$) respectively. This results in the heat energy generation (\ref{entangle}) due to the variation of the entropy density  of the null surface identified as $\delta_{\lambda (l)} s$, where $s$ is again proportional to $\sqrt{q}$ of the area element of $S_t$. The  irreversible heat generation (\ref{irrev}) is due to the viscous dissipative effects present in the null surface. We also identified that this dissipative heat generation must be entirely due to geometric fluxes since the dissipation term contains only geometrical quantities established on $\mathcal{H}$. Finally, we showed for the explicit case of a stationary black hole system, that the integrated form of the NRC (for $l^a$) over the virtual displacement produces the generalized Smarr formula (\ref{B11}). We also showed that the energy term (in both our interpretations) is proportional to the Komar energy term (\ref{B8}) for this special case.  

Next we focused on the more relevant component $R_{ab} l^a k^b$ for providing the thermodynamical interpretation in a covariant fashion. In literature, previous works \cite{Padmanabhan:2002sha, Kothawala:2007em, Chakraborty:2015aja} had solidified the fact that $R_{ab}l^a k^b$ can be provided a thermodynamical relationship which is structurally quite similar to the first law of thermodynamics. However they had been proposed in an adapted coordinate system called the GNC system. This results in the expression of the energy being dependant on the GNC coordinates. Here, we have tried to show, whether a similar interpretation can be provided without the need of adapting any coordinate system w.r.t the null surface. In our approach we started out with the NRC equation (for $k^a$ field) (\ref{thermoevolution_trace}) and provided a virtual displacement $\delta \lambda_{(k)}$. We then integrated the resulting equation over the $2$-surface $S_t$. This procedure allowed us to obtain our required thermodynamic interpretation in a covariant fashion. However our proposed interpretation does have major differences with the previous approaches. We have not invoked in our analysis any specific gravitational field equations and hence proposed that our interpretation is not specific to any particular theory of gravity. This required us to propose that the entropy density of the null surface is actually the entanglement entropy density assigned to $\mathcal{H}$ by a null observer moving along the integral curves of the null generators $l^a$. This is because in our case the entropy density is actually proportional to $\sqrt{q}$ of our area element on the $2$-surface $S_t$. The temperature is again found to be proportional to the non-affinity parameter $\kappa$ associated with the null generators $l^a$. To have a consistent thermodynamic interpretation irrespective of any particular theory of gravity we have defined a so called geometric pressure $P = -1/(8 \pi G)(R_{ij}l^i k^j+\frac{1}{2} R)$. This is in contrast with the earlier methods defining the pressure entirely through the matter energy tensor. Moreover, the identified energy here is in covariant form and so can be applied to any metric adapted to the null surface. This added advantage of our formalism must be very useful for further progress of this field.

In this present work we have considered only a Riemannian manifold with the Levi-Civita connection i.e our spacetime geometry has no torsion. An interesting perspective to look at is to see if analogous physical/thermodynamic interpretations can be provided to $\bar{R}_{ab}l^a l^b$, $\bar{R}_{ab}l^a k^b$ and $\bar{R}_{ab}l^a q^b_{~c}$, where $\bar{R}_{ab}$ is the Ricci tensor corresponding to spacetimes with torsion. The idea of Jacobson of deriving the gravitational field equations from a local constitutive relation \cite{Jacobson:1995ab} has been extended to case of spacetimes with torsion in \cite{Dey:2017fld}. The equilibrium Clausius relation has been replaced by the entropy balance law to incorporate for the internal irreversible entropy generation term. The authors of \cite{Dey:2017fld} obtain the Einstein-Cartan-Kibble-Sciama gravitational field equations under such a procedure and identify the corresponding internal entropy generation term. It would be a fruitful exercise to carry out our procedure of taking the NRC for $l^a$ along with the virtual displacement $\delta \lambda_{(l)}$ for the case of spacetimes with torsion and provide an analogous thermodynamic interpretation. However the terms like energy and dissipation need to be properly understood in this context. Similarly it would be quite instructive to work out the NRC corresponding to the auxiliary null vector field $k^a$ under the inclusion of torsion. Corresponding attachment of a thermodynamic interpretation under a virtual displacement $\delta \lambda_{(k)}$ needs to be analysed carefully by understanding the nature of the energy and geometric work function. Similarly seeing what modifications arise to the DNS equation under the consideration of $\bar{R}_{ab}l^a q^b_{~c}$ is quite interesting. We certainly aim towards such directions and hope to report it in future.

Finally, we feel that the present results are very generic to any null surface as the underlying dynamics of background has not been explicitly used. Instead the construction is purely based on the geometrical properties of the aforesaid surface.
We hope that this new foray into the the well known study of the thermodynamical structure of a generic null hypersurface via our covariant approach will help to shed some light on it.
%%%%%%%%%%%%%%%%%%%%%%%%%%%%%%%%%%%%%%%%%%%%%%%%%%%%%
\vskip 3mm
\noindent
{\bf Acknowledgement:} {\sc This work is dedicated to those who are helping us to fight against COVID-19 across the globe.}
%%%%%%%%%%%%%%%%%%%%%%%%%%%%%%%%%%%%%%%%%%%%%%%%%%%

\vskip 4mm
\appendix
\section{Derivation of Eq. (\ref{thermoevolution_trace})}\label{App1}
Upon taking the trace on both sides of the equation (\ref{transverse_evol}), let us now concentrate on the LHS first.
The irreducible decomposition of the transversal deformation rate tensor is given by,
\begin{equation}
\Xi_{ab} = \frac{1}{2}q_{ab} \theta_{(k)} + \sigma_{(k)_{ab}} ~,
\label{irrep}
\end{equation}
where $\theta_{(k)}$ is the trace part and $\sigma_{(k)_{ab}}$ is the traceless symmetric part.
Therefore, the trace of  LHS of (\ref{transverse_evol}) can be expressed as
\begin{equation}
	q^{ij}\pounds_l \Xi_{ij} = \theta_{(l)} \theta_{(k)} + \pounds_l \theta_{(k)} + q^{ij}\pounds_l \sigma_{(k)_{ij}} ~.
\label{c15}
\end{equation}
Focussing on the term $q^{ij}\pounds_l \sigma_{(k)_{ij}}$ in (\ref{c15}), and using the fact that 
\begin{equation}
\nabla_a l_b = \Theta_{ab} + \omega_a l_b - l_a(k^i \nabla_i l_b) ~,
\label{delalb}
\end{equation} 
and $\sigma_{(k)_{ij}}$ being orthogonal to $l^i$, we have,
\begin{eqnarray}
q^{ij}\pounds_l \sigma_{(k)_{ij}} = 2 \sigma_{(k)_{ab}} \sigma^{ab} ~,
\end{eqnarray} 
where $\sigma^{ab}$ is the shear tensor associated with the null generators $l^a$ of the null hypersurface $\mathcal{H}$. As a result the LHS of trace of Eq. (\ref{transverse_evol}) yields,
\begin{equation}
q^{ij}\pounds_l \Xi_{ij} = \theta_{(l)} \theta_{(k)} + l^i \nabla_i \theta_{(k)} + 2 \sigma_{(k)_{ab}} \sigma^{ab}~.
\label{traceLHS1} 
\end{equation}

Having done this, we now focus on trace of the RHS of Eq. (\ref{transverse_evol}). This is given by
\begin{eqnarray}
{^2\mathcal{D}_a }\Omega^a + \Omega_a \Omega^a -\frac{1}{2}{^2 R} + \frac{1}{2}q^{ij} R_{ij} 
\nonumber \\
-\Big(\kappa + \frac{\theta_{(l)}}{2} \Big) \theta_{(k)}
- \frac{1}{2} \theta_{(l)} \theta_{(k)} + 2 \Theta_{ab} \Xi^{ab}\nonumber ~.
 \label{traceRHS1}
\end{eqnarray}
Upon using the irreducible decomposition of both $\Theta_{ab}$ = $(1/2) q_{ab} \theta + \sigma_{ab}$ and $\Xi_{ab}$ (see Eq. (\ref{irrep})) we obtain the RHS of Eq. (\ref{transverse_evol}) as,
\begin{eqnarray}
{^2\mathcal{D}_a }\Omega^a + \Omega_a \Omega^a -\frac{1}{2}{^2 R} + \frac{1}{2}q^{ij} R_{ij} 
\nonumber
\\
- \kappa \theta_{(k)} + 2 \sigma_{ab} \sigma_{(k)}^{ab} ~.
\label{traceRHS2}
\end{eqnarray}
Note that the term $\frac{1}{2} q^{ij} R_{ij}$ is equal to $\frac{1}{2} R + R_{ij} l^i k^j$. Using this, (\ref{traceRHS2}) goes over to,
\begin{eqnarray}
{^2\mathcal{D}_a }\Omega^a + \Omega_a \Omega^a -\frac{1}{2}{^2 R} + \frac{1}{2} R + R_{ij} l^i k^j \nonumber \\
- \kappa \theta_{(k)} + 2 \sigma_{ab} \sigma_{(k)}^{ab} ~.
\label{traceRHS3}
\end{eqnarray}
%Finally using the Einstein's field equations the term $R_{ij} l^i k^j$ can be expressed as,
%\begin{equation}
%R_{ij} l^i k^j = 8 \pi T_{ij} l^i k^j - \frac{1}{2} R ~,
%\label{C22}
%\end{equation}
%which when implemented in (\ref{traceRHS3}) yields, for the trace of the RHS of Eq. (\ref{transverse_evol}) the following,
%\begin{eqnarray}
%{^2\mathcal{D}_a }\Omega^a + \Omega_a \Omega^a -\frac{1}{2}{^2 R} + 8 \pi T_{ij} l^i k^j \nonumber \\
% - \kappa \theta_{(k)} + 2 \sigma_{ab} \sigma_{(k)}^{ab} %~.
%\label{traceRHS4}
%\end{eqnarray}
Finally equating (\ref{traceLHS1}) and (\ref{traceRHS3}) we obtain the identity (\ref{thermoevolution_trace}).

\section{Derivation of Eq. (\ref{eqn20})}\label{proof}
Before delving into the derivation, we note two relations involving the covariant derivatives of the null normals $l^a$ and $k^a$, which we are going to put to heavy usage,
\begin{equation}
\nabla_{a} l_b = \Theta_{ab} + \omega_a l_b - l_a (k^i \nabla_i l_b)~,
\label{eqnb1}
\end{equation}
and
\begin{equation}
\nabla_{a} k_b = \Xi_{ab} - \Omega_a k_b -k_a \omega_b - l_a(k^i \nabla_i k_b) ~.
\label{eqnb2}
\end{equation}
We start with the Ricci identity for the null normals $l^a$ and $k^a$,
\begin{eqnarray}
l^a(\nabla_a \nabla_b k_c) = l^a(\nabla_b \nabla_a k_c) - R_{abfc}l^a k^f ~.
\label{eqnb3}
\end{eqnarray}
We focus on the LHS of Eq.  (\ref{eqnb3}). Upon using Eq. (\ref{eqnb2}), we obtain,
\begin{eqnarray}
l^a (\nabla_a \nabla_b k_c) &=& l^a(\nabla_a \Xi_{bc}) - (l^a \nabla_a \Omega_b)k_c - \Omega_b(l^a \nabla_a k_c) 
\nonumber 
\\
&-& (l^a \nabla_a k_b) \omega_c - k_b (l^a \nabla_a \omega_c)  \nonumber \\
& -& (l^a \nabla_a l_b)(k^i \nabla_i k_c) - l_b l^a \nabla_a(k^i \nabla_i k_c).
\label{eqnb4}
\end{eqnarray}
Upon using the relations $l^a \nabla_a l_b = \kappa l_b$ , $l^a \nabla_a k_b = \omega_b$ and $\omega_a = \Omega_a -\kappa k_a$, we contract the above Eq.  (\ref{eqnb4}) with $k^bl^c$ to have,
\begin{eqnarray}
l^a (\nabla_a \nabla_b k_c)k^bl^c =& -\Omega_b \Omega^b + \kappa l^c k^i (\nabla_i k_c)\nonumber \\
& + l^c l^a \nabla_a(k^i \nabla_i k_c) ~.
\label{eqnb5}
\end{eqnarray}
We now focus on the first term of the RHS of Eq.  (\ref{eqnb3}) i.e $ l^a \nabla_b (\nabla_a k_c)$. Again upon using Eq. (\ref{eqnb2}), we have,
\begin{eqnarray}
l^a \nabla_b (\nabla_a k_c) =& -\Xi_{ac}(\nabla_b l^a) + \Omega_a (\nabla_b l^a) k_c \nonumber \\
& -(l^a \nabla_b k_a) \omega_c + \nabla_b \omega_c ~.
\label{eqnb6}
\end{eqnarray}
Contracting the above Eq. (\ref{eqnb6}) with $k^b l^c$ and using the fact that $\Omega_a l^a = 0$, $\Omega_a k^a = 0$ and $\omega_a l^a = \kappa$ we obtain,
\begin{eqnarray}
l^a (\nabla_b \nabla_a k_c) k^b l^c &= -\Omega_a(\nabla_b l^a)k^b - \kappa k^b l^a (\nabla_b k_a)\nonumber \\
& + k^b l^c(\nabla_b \omega_c) \nonumber \\
& = l^a k^b(\nabla_b \Omega_a) - \kappa l^c k^i (\nabla_i k_c)\nonumber \\
& + k^b l^c(\nabla _b \omega_c) ~.
\label{eqnb7}
\end{eqnarray}
We focus on the first term of the RHS of Eq.  (\ref{eqnb7}) i.e $l^a k^b(\nabla_b \Omega_a)$,
\begin{eqnarray}
l^a k^b(\nabla _b \Omega_a) &= (q^{ab}-g^{ab}- k^a l^b) (\nabla_b \Omega_a) \nonumber \\
&= q^{ab}\Big(\delta_b^i \delta_a ^k (\nabla_i \Omega_k)\Big) - (\nabla_b \Omega^b) \nonumber \\
& + l^b \Omega^a (\nabla_b k_a) ~.
\label{eqnb8}
\end{eqnarray}
Upon using the completeness relation $\delta^a_b = q^a_b - l_b k^a - l^a k_b$, we have after some simple algebra, 
\begin{eqnarray}
l^a k^b(\nabla _b \Omega_a) = q^{ab}\Big( {^{2}\mathcal{D}_b \Omega_a}\Big) - (\nabla_a \Omega^a) + \Omega_a \Omega^a ~.
\label{eqnb9}
\end{eqnarray}
Putting Eq. (\ref{eqnb9}) into Eq. (\ref{eqnb7}), we obtain, 
\begin{eqnarray}
l^a (\nabla_b \nabla_a k_c) k^b l^c &= q^{ab}\Big( {^{2}\mathcal{D}_b \Omega_a}\Big) - (\nabla_a \Omega^a)\nonumber \\
& + \Omega_a \Omega^a - \kappa l^c k^i (\nabla_i k_c) \nonumber \\
&+ k^b l^c(\nabla _b \omega_c) ~.
\label{eqnb10}
\end{eqnarray}
We now contract the Ricci Identity i.e Eq. (\ref{eqnb3}) on both sides with $k^b l^c$. Following this we use the relations (\ref{eqnb5}) and (\ref{eqnb10}) onto the contracted Ricci Identity to obtain,
\begin{equation}
\begin{split}\label{eqnb11}
R_{abcd}l^a k^b k^c l^d &= q^{ab}\Big( {^{2}\mathcal{D}_b \Omega_a}\Big) - (\nabla_a \Omega^a)  \\
& + 2\Omega_a \Omega^a - 2\kappa l^c k^i (\nabla_i k_c) \\
& + k^b l^c(\nabla _b \omega_c) - l^c l^a \nabla_a(k^i \nabla_i k_c) ~.
\end{split}
\end{equation} 
The term $- 2\kappa l^c k^i (\nabla_i k_c)$ can further be manipulated as,
\begin{equation}
\begin{split}
- 2\kappa l^c k^i (\nabla_i k_c) &= -2\kappa(q^{ci} - g^{ci}- k^cl^i) (\nabla_i k_c)  \\
&= -2\kappa \theta_{(k)} + 2\kappa(\nabla_a k^a) ~.
\label{eqnb12}
\end{split}
\end{equation}
Putting Eq.  (\ref{eqnb12}) in Eq.  (\ref{eqnb11}) we obtain as a result,
\begin{equation}\label{eqnb13}
\begin{split}
R_{abcd}l^a k^b k^c l^d =& {^{2}\mathcal{D}_b \Omega^b} - (\nabla_a \Omega^a) + 2 \Omega_a\Omega^a \\
& - 2\kappa \theta_{(k)} + 2\kappa (\nabla_a k^a) + k^b l^c(\nabla_b \omega_c) \\
& - l^c l^a \nabla_a (k^i \nabla_i k_c) ~.
\end{split}
\end{equation}
Following this result, we focus on the last term on the RHS of Eq.  (\ref{eqnb13}) i.e $l^c l^a \nabla_a (k^i \nabla_i k_c)$ and manipulate it in the following sense,
\begin{equation}
\begin{split}
l^c l^a \nabla_a (k^i \nabla_i k_c) = l^c \omega^i (\nabla_i k_c) + l^c l^a k^i \nabla_a(\nabla_i k_c) \nonumber ~.
\end{split}
\end{equation}
Upon using Eq.  (\ref{eqnb2}), we have,
\begin{equation}\label{eqnb14}
\begin{split}
l^c l^a \nabla_a (k^i \nabla_i k_c) = \Omega_i \Omega^i - \kappa l^c(k^i \nabla_i k_c) + l^cl^ak^i \nabla_a(\nabla_i k_c)~.
\end{split}
\end{equation}
We proceed to manipulate the last term on the RHS of Eq.  (\ref{eqnb14}) with the help of Eq.  (\ref{eqnb2}),
\begin{equation}\label{eqnb15}
\begin{split}
&l^cl^ak^i \nabla_a(\nabla_i k_c) =l^a (q^{ic} - g^{ic} - l^i k^c)\nabla_a (\nabla_i k_c) \\
& = q^{ic} l^a \nabla_a \Xi_{ic} - l^a q^{ic} \Omega_i (\nabla_a k_c) - l^a q^{ic}(\nabla_a k_i)\omega_c \\
& - l^a \nabla_a (\nabla_i k^i) - l^c k^a(\nabla_c \omega_a) ~.
\end{split}
\end{equation}
Putting Eq.  (\ref{eqnb15}) in Eq.  (\ref{eqnb14}) we obtain along with the use of Eq.  (\ref{eqnb2}),
\begin{equation}\label{eqnb16}
\begin{split}
l^c l^a \nabla_a (k^i \nabla_i k_c) &= \Omega_i \Omega^i - \kappa l^c(k^f \nabla_f k_c) + q^{ic}(l^a \nabla_a \Xi_{ic}) \\
& - 2l^a \Omega^c(\nabla_a k_c) - l^a \nabla_a(\nabla_i k^i) - l^c k^a(\nabla_c \omega_a)  \\
& = -\Omega_i \Omega^i - \kappa l^c(k^f \nabla_f k_c) + q^{ic}(l^a \nabla_a \Xi_{ic}) \\
& - l^a \nabla_a(\nabla_i k^i) - l^c k^a (\nabla_c \omega_a) ~.
\end{split}
\end{equation}
Putting the value of $l^c l^a \nabla_a (k^i \nabla_i k_c)$ from Eq.  (\ref{eqnb16}) in Eq. (\ref{eqnb13}), we obtain,
\begin{equation}
\begin{split}
R_{abcd}l^a k^b k^c l^d =& {^{2}\mathcal{D}_b \Omega^b} - (\nabla_a \Omega^a) + 3 \Omega_a \Omega^a -\kappa l^c k^i(\nabla_i k_c) \\
& - q^{ab} (l^i \nabla_i \Xi_{ab}) + l^a \nabla_a(\nabla_i k^i) \\
& + (l^ck^b + k^cl^b) (\nabla_c \omega_b) \\
& = {^{2}\mathcal{D}_a \Omega^a} - (\nabla_a \Omega^a) + 3 \Omega_a \Omega^a -\kappa l^c k^i(\nabla_i k_c) \\
& - q^{ab} (l^i \nabla_i \Xi_{ab}) + l^a \nabla_a(\nabla_i k^i)\\
& + (q^{cb}-g^{cb})(\nabla_c \omega_b) ~.
\end{split}
\end{equation} 
Expanding the above result and after a few lines of simple manipulations we obtain,
\begin{equation}\label{eqnb18}
\begin{split}
R_{abcd}l^a k^b k^c l^d =&{^{2}\mathcal{D}_a \Omega^a} - (\nabla_a \Omega^a) + 3 \Omega_a \Omega^a -\kappa l^c k^i(\nabla_i k_c) \\
& - q^{ab} (l^i \nabla_i \Xi_{ab}) + q^{ab}(\nabla_a \omega_b)\\
& - (\nabla_a l^i)(\nabla_i k^a) - R_{ab}l^a k^b ~.
\end{split}
\end{equation}
To this end, we focus at the $- q^{ab} (l^i \nabla_i \Xi_{ab})$ term and using the fact,
\begin{equation}
\pounds_{l} \Xi_{ab} = l^i\nabla_i \Xi_{ab}+ \Xi_{ai}(\nabla_b l^i) + \Xi_{ib}(\nabla_a l^i) \nonumber
\end{equation}
along with Eq.  (\ref{eqnb1}), we have,
\begin{equation}\label{eqnb19}
\begin{split}
- q^{ab} (l^i \nabla_i \Xi_{ab}) &= -q^{ab} \pounds_{l} \Xi_{ab} + 2 \Xi_{ab} \Theta^{ab} \\
& = -q^{ab} \pounds_{l} \Xi_{ab} + \theta_{(l)}\theta_{(k)} + 2 \sigma_{ab} \sigma_{(k)}^{ab} ~.
\end{split}
\end{equation}
Upon using the irreducible decomposition of the transversal deformation rate tensor $\Xi_{ab}$, it is fairly straightforward to show that,
\begin{equation}\label{eqnb20}
q^{ab} \pounds_{l} \Xi_{ab} = \theta_{(l)}\theta_{(k)} + 2 \sigma_{ab} \sigma_{(k)}^{ab} + l^i \nabla_i \theta_{(k)}  ~.
\end{equation} 
Using Eq. (\ref{eqnb20}) in Eq. (\ref{eqnb19}), we obtain,
\begin{equation}\label{eqnb21}
- q^{ab} (l^i \nabla_i \Xi_{ab}) = - l^i \nabla_i \theta_{(k)} ~.
\end{equation}
Upon using Eq. (\ref{eqnb21}) and the relation $-\kappa l^c k^i(\nabla_i k_c) = -\kappa \theta_{(k)} + \kappa(\nabla_a k^a)$ in Eq. (\ref{eqnb18}), we have as a result,
\begin{equation}\label{eqnb22}
\begin{split}
R_{abcd}l^a k^b k^c l^d =&{^{2}\mathcal{D}_a \Omega^a} - (\nabla_a \Omega^a) + 3 \Omega_a \Omega^a -\kappa \theta_{(k)}\\
& +\kappa(\nabla_a k^a) - l^i \nabla_i \theta_{(k)} + q^{ab}(\nabla_a \omega_b)\\
&-(\nabla_a l^i)(\nabla_i k^a) - R_{ab}l^a k^b ~.
\end{split}
\end{equation} 
Let us now take a look at the term $(\nabla_a l^i)(\nabla_i k^a) $. Using the relations (\ref{eqnb1}) and (\ref{eqnb2}), it can manipulated quite simply to be,
\begin{equation}\label{eqnb23}
(\nabla_a l^i)(\nabla_i k^a) = \Theta_{ab} \Xi^{ba} + \Omega_a \Omega^a - k^a \nabla_a \kappa + l^a k^b (\nabla_b \omega_a) ~.
\end{equation}
Looking at the last term on the RHS of Eq.  (\ref{eqnb23}), i.e $l^a k^b (\nabla_b \omega_a)$ , we obtain in the process of manipulation as 
\begin{equation}\label{eqnb24}
\begin{split}
l^a k^b (\nabla_b \omega_a) &= \Big(q^{ab}-g^{ab}-l^b k^a\Big)(\nabla_b \omega_a) \\
& = q^{ab}(\nabla_b \omega_a) - (\nabla_b \omega^b) + \Omega_a \Omega^a ~.
\end{split}
\end{equation}
Equating Eq. (\ref{eqnb24}) in Eq.  (\ref{eqnb23}), we have,
\begin{equation}\label{eqnb25}
\begin{split}
(\nabla_a l^i)(\nabla_i k^a) - q^{ab}(\nabla_b \omega_a) &= \Theta_{ab} \Xi^{ba} + 2 \Omega_a \Omega^a \\
& - k^a \nabla_a \kappa  -(\nabla_b \omega^b) ~.
\end{split}
\end{equation} 
Looking at Eq. (\ref{eqnb22}) we manipulate the terms $(\nabla_a \Omega^a) - \kappa (\nabla_a k^a)$ using the relation $\omega_a = \Omega_a - \kappa k_a$,
\begin{equation}\label{eqnb26}
(\nabla_a \Omega^a) - \kappa (\nabla_a k^a) = (\nabla_a \omega^a) + (k^a \nabla_a \kappa) ~.
\end{equation}
To this end, we obtain, using Eq. (\ref{eqnb26}) and Eq.  (\ref{eqnb25}),
\begin{equation}\label{eqnb27}
\begin{split}
-(\nabla_a \Omega^a) + \kappa (\nabla_a k^a) +q^{ab}(\nabla_a \omega_b) -(\nabla_a l^i)(\nabla_i k^a) = \\
-\Theta_{ab}\Xi^{ab} -2 \Omega_a \Omega^a =\\
-\frac{1}{2} \theta_{(l)}\theta_{(k)} -\sigma_{ab} \sigma_{(k)}^{ab} - 2\Omega_a \Omega^a ~.
\end{split}
\end{equation} 
Finally, putting Eq. (\ref{eqnb27}) in Eq. (\ref{eqnb22}) we obtain our desired result,
\begin{equation}
\begin{split}
R_{abcd}l^a k^b k^c l^d &= {^{2}\mathcal{D}_a \Omega^a} + \Omega_a \Omega^a - \kappa \theta_{(k)} - l^i \nabla_i \theta_{(k)} \nonumber \\
& -\frac{1}{2}\theta_{(l)} \theta_{(k)} - \sigma_{ab}\sigma_{(k)}^{ab} - R_{ab}l^a k^b ~.
\end{split}
\end{equation}
%%%%%%%%%%%%%%%%%%%%%%%%%%%%%%%%%%%%%%%%%%%%%%%%%%%%%%%%%%%%%%%%%%%%%%
\section{Derivation of Eq. (\ref{d15})}{\label{totalderq}}
We have by definition,
\begin{eqnarray}
\Xi_{ab} = \frac{1}{2} q_a^{~c}q_b^{~d} \pounds_{k} q_{cd} ~.
\end{eqnarray}
Taking trace of the above equation, we obtain, and using the irreducible decomposition of $\Xi_{ab}$, i.e $\Xi^{ab} = 1/2  q^{ab} \theta_{(k)} + \sigma_{(k)}^{ab}$ we have
\begin{eqnarray}
\theta_{(k)} = \frac{1}{2} q^{ab} \pounds_{k} q_{ab}~.
\end{eqnarray}
Using the definition of the projection tensor and the Lie derivative its quite simple to show,
\begin{equation}
\theta_{(k)}  = q^{ij}(\nabla_i k_j) = \frac{1}{2} q^{ij}\pounds_{k} q_{ij} = \frac{1}{\sqrt{q}}\pounds_k \sqrt{q} ~.
\end{equation}
Now we show that $\theta_{(k)}$ is actually equivalent to $-\frac{1}{\sqrt{q}}\frac{d}{d \lambda_{(k)}} \sqrt{q}$. For that we note that the basis vectors $e^a_{~A}$ are actually Lie transported along the $k^a$ fields i.e $\pounds_{k}e^a_{~A} =0$. As a result we can write,
\begin{eqnarray}
\frac{d \sqrt{q}}{d \lambda_{(k)}} &=& \frac{1}{2} \sqrt{q} q^{AB} \frac{d}{d \lambda_{(k)}} q_{AB} = \frac{1}{2} \sqrt{q} q^{AB} \frac{d}{d \lambda_{(k)}}(g_{ab}e^a_{~A} e^b_{~B}) \nonumber \\
&=& -\frac{1}{2} \sqrt{q} q^{AB} k^i \nabla_i (g_{ab} e^a_{~A} e^b_{~B} ) ~.
\end{eqnarray}
Now using the fact that $k^i \nabla_i e^a_{~A} = e^i_{~A} \nabla_i k^a$, we have,
\begin{eqnarray}
\frac{d \sqrt{q}}{d \lambda_{(k)}} &=& -\frac{1}{2} \sqrt{q} q^{AB}  \Big(e^a_{~A}e^i_{B} \nabla_i k_a + e^b_{~B} e^i_{~A} \nabla_i k_b \Big) \nonumber \\
&=&  -\sqrt{q} q^{ab} (\nabla_a k_b) ~.
\end{eqnarray}
Hence, the result of (\ref{d15}) follows.

\section{Derivation of  $l^i \nabla_i \theta_{(k)} = k^i \nabla_i \theta_{(l)}$}
\label{App2}
Here we sketch an outline of the proof of $l^i \nabla_i \theta_{(k)} = k^i \nabla_i \theta_{(l)}$,
\begin{eqnarray}
&l^i \nabla_i\theta_{(k)} = \frac{d}{d \lambda_{(l)}}\Big(-\frac{1}{\sqrt{q}}\frac{d}{d \lambda_{(k)}}\sqrt{q}\Big) \nonumber \\
 &= \frac{1}{\sqrt{q}^2} \Big(\frac{d}{d \lambda_{(k)}}\sqrt{q}\Big) \Big(\frac{d}{d \lambda_{(l)}\sqrt{q}}\Big) - \frac{1}{\sqrt{q}}\Big(\frac{d}{d \lambda_{(k)}}\frac{d}{d \lambda_{(l)}}\sqrt{q}\Big) \nonumber \\
 & = \frac{d}{d \lambda_{(k)}}\Big(-\frac{1}{\sqrt{q}}\Big)\frac{d}{d \lambda_{(l)}}\sqrt{q} + \Big(-\frac{1}{\sqrt{q}}\Big)\frac{d}{d \lambda_{(k)}} \Big(\frac{d}{d \lambda_{(l)}} \sqrt{q}\Big) \nonumber \\
& = -\frac{d}{d \lambda_{(k)}}\Big(\frac{1}{\sqrt{q}}\frac{d}{d \lambda_{(l)}}\sqrt{q}\Big)  = k^i \nabla_i \theta_{(l)}
\end{eqnarray}

\section{Connection with existing results}\label{existing_results}
In the section \ref{rablakb} we landed ourselves with a covariant expression of the energy (\ref{energy111}) of the null surface $S_t$ associated with a virtual displacement $\delta \lambda_{k}$ in the outgoing auxiliary null direction. We now aim to compute this expression of the energy in the GNC system. To this end, we mention that the metric expressed in the GNC $(u,r,x^A)$ reads,
\begin{equation}\label{GNC_metric}
ds^2 = -2r\alpha du^2 + 2 du dr -2 r \beta_A du dx^A + q_{AB} dx^A dx^B ~,
\end{equation} 
where the six independent parameters $(\alpha, \beta_A, q_{AB})$ are dependant on the coordinates $(u,r,x^A)$. The null hypersurface in this system is stationed at $r=0$. The relevant inverse metric as well as its Christoffel connection coefficients have been calculated in \cite{Parattu:2015gga}. 
The components of the null normal and the auxiliary null normal in this coordinate system are,
\begin{eqnarray}\label{c38}
\begin{split}
l_a &= (0,1,0,0) \quad \quad k_a = (-1,0,0,0)\\
&l^a = (1, 2r \alpha + r^2 \beta^2, r\beta^A) \quad k^a = (0,-1,0,0) ~.
\end{split}
\end{eqnarray}
Before proceeding ahead, we now invoke the Einstein's field equations and note that the work function previously defined as $P = -1/(8 \pi G)(R_{ij}l^i k^j+\frac{1}{2} R)$ when evaluated on the null hypersurface $r =0$, yields $ P = -1/(8 \pi G)(R_{ij}l^i k^j+\frac{1}{2} R)= (-T_{ij} l^i k^j) = (-T^{ij} l_i k_j) = -T^a_{~b} l_a k^b = T^{ur} = T^r_{~r}$ = $T_{ur}  = T^u_{~u}$. In static spherically symmetric spacetimes $T^r_{~r}$ has the interpretation of being the radial or the normal pressure \cite{Hayward:1997jp, Kothawala:2010bf}. Hence the integral of the work function $F = \int_{S_t} d^2 x \sqrt{q} P$ in the static spherically symmetric case is $F = \int_{S_t} d^2 x \sqrt{q} P = \int_{S_t} d^2 x \sqrt{q} T^r_{~r}$, which is to be interpreted as the average normal force on $S_t$ 

We note that all the quantities in the integrand of the expression of energy (\ref{energy111}) are to be evaluated on the null hypersurface i.e at $r=0$. 
Looking at the term $\theta_{(l)}$, we obtain, 
\begin{equation}
\theta_{(l)} = q^{ab}\nabla_a l_b = -q^{AB}\Gamma^{r}_{AB} ~.
\end{equation} 
The value of $\Gamma^{r}_{AB}$ ,
\begin{equation}
\begin{split}
\Gamma^{r}_{AB} &= -\frac{1}{2}\partial_u q_{AB} -\frac{1}{2} (r^2\beta^2+ 2r\alpha)\partial_r q_{AB}\\
& + \frac{1}{4} r \Big({^2\mathcal{D}_A \beta_B}+{^2\mathcal{D}_B \beta_A}\Big) ~.
\end{split}
\end{equation}
Evaluating $\theta_{(l)}$ on the null hypersurface,
\begin{equation}
\begin{split}\label{c41}
\theta_{(l)|_{r=0}} &= -q^{AB}\Gamma^{r}_{AB |_{r=0}} = \frac{1}{2} q^{AB}\partial_u q_{AB} \\
& = \frac{1}{\sqrt{q}}\Big(\partial_u \sqrt{q}\Big) = \partial_u \Big(\ln \sqrt{q}\Big) ~. 
\end{split}
\end{equation}
Computation of $k^i \nabla_i \theta_{(l)}$ on the null hypersurface, with the components of $k^i$ given in (\ref{c38}), yields,
\begin{equation}\label{e28}
k^i \nabla_i \theta_{(l)|_{r=0}} = -\partial_r\Big(\frac{1}{\sqrt{q}}\partial_u \sqrt{q}\Big) ~.
\end{equation}
Looking at the computation of $\theta_{(k)}$, we have,
\begin{equation}
\theta_{(k)} = q^{ab}\nabla_a k_b = q^{AB}\Gamma^{u}_{AB} ~.
\end{equation}
The value of  $\Gamma^{u}_{AB}$ is $-\frac{1}{2}\partial_r q_{AB}$. Evaluating $\theta_{(k)}$ on the null hypersurface,
\begin{equation}\label{c44}
\theta_{(k)|_{r=0}} = q^{AB}\Gamma^{u}_{AB|_{r=0}} = -\frac{1}{\sqrt{q}}\Big(\partial_r \sqrt{q}\Big) ~.
\end{equation}
This allows us to have,
\begin{equation}\label{e30}
\begin{split}
\theta_{(l)} \theta_{(k)|_{r=0}} + k^i \nabla_i \theta_{(l)|_{r=0}} = -\frac{1}{\sqrt{q}}\partial_r \partial_u \sqrt{q} ~.
\end{split}
\end{equation}
Next, noting that $\Omega_a = \omega_a + \kappa k_a $, we have $\Omega_a \Omega^a = \omega_a \omega^a $, where $\omega_a$ refers to the rotation one form defined in the manifold. For the evaluation of $\omega_a$ = $l^i \nabla_i k_a$, with $l^i$ and $k_a$  provided from (\ref{c38}), we have,
\begin{equation}
\omega^u = 0 \quad \omega_r = 0 ~.
\end{equation}
As a consequence of this, we have $\omega_a \omega^a = \omega_A \omega^A$.
The relevant quantities are,
\begin{equation}
\begin{split}
\omega_A  &= \frac{1}{2}\beta_A + \frac{1}{2} r \Big(\partial_r \beta_A\Big) - \frac{1}{2} r \beta^B\Big(\partial_r q_{AB}\Big) \\
\omega^A &= q^{AC}\Big[\frac{1}{2}\beta_C + \frac{1}{2} r \Big(\partial_r \beta_C\Big) - \frac{1}{2} r \beta^B\Big(\partial_r q_{BC}\Big)\Big]
\end{split}
\end{equation}
Evaluation of $\Omega_a \Omega^a$ on the null hypersurface at $r=0$ yields,
\begin{equation}\label{e32}
\Omega_a \Omega^a_{|{r=0}} = \omega_A \omega^A_{|{r=0}} = \frac{1}{4}\beta_A \beta^A ~.
\end{equation}
Finally we are left with the evaluation of ${^2\mathcal{D}}_A \Omega^A$ on the null hypersurface. The calculations follow as,
\begin{eqnarray}
\Omega^A &=& q^{AB} \omega_B \nonumber \\
&=& q^{AB}\Big(\frac{1}{2}\beta_B + \frac{1}{2} r \Big(\partial_r \beta_B\Big) - \frac{1}{2} r \beta^D\Big(\partial_r q_{BD}\Big)\Big) 
\end{eqnarray} 
\begin{eqnarray}
{^2\mathcal{D}}_A \Omega^A &&= \frac{1}{\sqrt{q}}\partial_A \Big[\sqrt{q}\Big(\frac{1}{2}\beta^A + \frac{1}{2} r q^{AB}\Big(\partial_r \beta_B\Big) \nonumber \\
&& - \frac{1}{2} r q^{AB} \beta^D\Big(\partial_r q_{BD}\Big)\Big)\Big]
\end{eqnarray}
\begin{eqnarray}
{^2\mathcal{D}}_A \Omega^A_{|{r=0}} = \frac{1}{2}\frac{1}{\sqrt{q}} \partial_A(\sqrt{q}\beta^A) ~.
\label{c53}
\end{eqnarray}
In the G.N.C coordinates, the virtual displacement $\delta x^a = \frac{dx^a}{d\lambda_{(k)}}\delta \lambda = -k^a \delta \lambda =(0, \delta \lambda = \delta r,0,0)$.
Finally putting the values of the relevant quantities obtained  in (\ref{e30}), (\ref{e32}) and (\ref{c53}) into the expression of the energy in (\ref{energy111}), we obtain,
\begin{equation}\label{energyGNC}
\begin{split}
E &= \frac{1}{2}\int dr \Big(\frac{\chi}{2 G}\Big) \\
& -\frac{1}{8 \pi G} \int  dr \Big[\int_{S_t} d^2 x \sqrt{q}\Big(\frac{1}{\sqrt{q}}\partial_r \partial_u \sqrt{q}\\
& + \frac{1}{4}\beta_A \beta^A\Big) +\frac{1}{2}\frac{1}{\sqrt{q}} \partial_A(\sqrt{q}\beta^A)\Big] ~.
\end{split}
\end{equation}
We note that,
\begin{equation}
\int^{r=2}_{r=1} dr \int_{S_t} d^2 x \sqrt{q}\Big(\frac{1}{\sqrt{q}}\partial_r \partial_u \sqrt{q}\Big) = \int_{S_t} d^2 x \partial_u \sqrt{q}|^{r=2}_{r=1} ~.
\end{equation}
With this, the energy of the null hypersurface manifests as,
\begin{equation}\label{energyGNC1}
\begin{split}
E &=  \frac{1}{2}\int dr \Big(\frac{\chi}{2 G}\Big) - \frac{1}{8 \pi G} \int_{S_t} d^2 x \partial_u \sqrt{q}\\
& -\frac{1}{16 \pi G} \int dr \int_{S_t} d^2 x \sqrt{q} \Big[\frac{1}{2}\beta_A \beta^A + \frac{1}{\sqrt{q}} \partial_A(\sqrt{q}\beta^A)\Big]
\end{split}
\end{equation}
We find that the expression of the Energy obtained in the GNC (\ref{energyGNC1}) via the covariant form of the expression of the energy (\ref{energy111}) matches with Eq. ($53$) in \cite{Chakraborty:2015aja}. Provided the $2$-dimension surface $S_t$ is compact, the above expression can be simplified to,
\begin{equation}\label{energyGNC2}
\begin{split}
E &=  \frac{1}{2}\int dr \Big(\frac{\chi}{2 G}\Big) - \frac{1}{8 \pi G} \int_{S_t} d^2 x \partial_u \sqrt{q}\\
& -\frac{1}{16 \pi G} \int dr \int_{S_t} d^2 x \sqrt{q} \Big[\frac{1}{2}\beta_A \beta^A\Big]~.
\end{split}
\end{equation}
The reason as to why (\ref{energyGNC2}) is called the energy term is because it provides the expression of the energy in quite a few well known cases. For a review of these specific cases please see \cite{Chakraborty:2015aja}. As an example, for the Schwarzchild metric the energy term (\ref{energyGNC2}) reduces to the mass.

\bibliographystyle{elsarticle-num}
\bibliography{references}

\begin{thebibliography}{10}
\expandafter\ifx\csname url\endcsname\relax
  \def\url#1{\texttt{#1}}\fi
\expandafter\ifx\csname urlprefix\endcsname\relax\def\urlprefix{URL }\fi
\expandafter\ifx\csname href\endcsname\relax
  \def\href#1#2{#2} \def\path#1{#1}\fi

\bibitem{Bekenstein:1973ur}
J.~D. Bekenstein, {Black holes and entropy}, Phys. Rev. D 7 (1973) 2333--2346.
\newblock \href {http://dx.doi.org/10.1103/PhysRevD.7.2333}
  {\path{doi:10.1103/PhysRevD.7.2333}}.

\bibitem{Bardeen:1973gs}
J.~M. Bardeen, B.~Carter, S.~Hawking, {The Four laws of black hole mechanics},
  Commun. Math. Phys. 31 (1973) 161--170.
\newblock \href {http://dx.doi.org/10.1007/BF01645742}
  {\path{doi:10.1007/BF01645742}}.

\bibitem{Hawking:1971vc}
S.~Hawking, {Black holes in general relativity}, Commun. Math. Phys. 25 (1972)
  152--166.
\newblock \href {http://dx.doi.org/10.1007/BF01877517}
  {\path{doi:10.1007/BF01877517}}.

\bibitem{Hawking:1976de}
S.~Hawking, {Black Holes and Thermodynamics}, Phys. Rev. D 13 (1976) 191--197.
\newblock \href {http://dx.doi.org/10.1103/PhysRevD.13.191}
  {\path{doi:10.1103/PhysRevD.13.191}}.

\bibitem{Davies:1978mf}
P.~Davies, {Thermodynamics of Black Holes}, Proc. Roy. Soc. Lond. A A353 (1977)
  499--521.
\newblock \href {http://dx.doi.org/10.1098/rspa.1977.0047}
  {\path{doi:10.1098/rspa.1977.0047}}.

\bibitem{Wald:1999vt}
R.~M. Wald, {The thermodynamics of black holes}, Living Rev. Rel. 4 (2001) 6.
\newblock \href {http://arxiv.org/abs/gr-qc/9912119}
  {\path{arXiv:gr-qc/9912119}}, \href {http://dx.doi.org/10.12942/lrr-2001-6}
  {\path{doi:10.12942/lrr-2001-6}}.

\bibitem{Carlip:2014pma}
S.~Carlip, {Black Hole Thermodynamics}, Int. J. Mod. Phys. D 23 (2014) 1430023.
\newblock \href {http://arxiv.org/abs/1410.1486} {\path{arXiv:1410.1486}},
  \href {http://dx.doi.org/10.1142/S0218271814300237}
  {\path{doi:10.1142/S0218271814300237}}.

\bibitem{Wall:2018ydq}
A.~C. Wall, {A Survey of Black Hole Thermodynamics.~~}\href
  {http://arxiv.org/abs/1804.10610} {\path{arXiv:1804.10610}}.

\bibitem{Parattu:2013gwa}
K.~Parattu, B.~R. Majhi, T.~Padmanabhan, {Structure of the gravitational action
  and its relation with horizon thermodynamics and emergent gravity paradigm},
  Phys. Rev. D 87~(12) (2013) 124011.
\newblock \href {http://arxiv.org/abs/1303.1535} {\path{arXiv:1303.1535}},
  \href {http://dx.doi.org/10.1103/PhysRevD.87.124011}
  {\path{doi:10.1103/PhysRevD.87.124011}}.

\bibitem{Chakraborty:2016dwb}
S.~Chakraborty, S.~Bhattacharya, T.~Padmanabhan, {Entropy of a generic null
  surface from its associated Virasoro algebra}, Phys. Lett. B 763 (2016)
  347--351.
\newblock \href {http://arxiv.org/abs/1605.06988} {\path{arXiv:1605.06988}},
  \href {http://dx.doi.org/10.1016/j.physletb.2016.10.059}
  {\path{doi:10.1016/j.physletb.2016.10.059}}.

\bibitem{Bhattacharya:2018epn}
K.~Bhattacharya, B.~R. Majhi, {Noncommutative Heisenberg algebra in the
  neighbourhood of a generic null surface}, Nucl. Phys. B 934 (2018) 557--577.
\newblock \href {http://arxiv.org/abs/1802.02862} {\path{arXiv:1802.02862}},
  \href {http://dx.doi.org/10.1016/j.nuclphysb.2018.07.025}
  {\path{doi:10.1016/j.nuclphysb.2018.07.025}}.

\bibitem{Maitra:2018saa}
M.~Maitra, D.~Maity, B.~R. Majhi, {Symmetries near a generic charged null
  surface and associated algebra: an off-shell analysis}, Phys. Rev. D 97~(12)
  (2018) 124065.
\newblock \href {http://arxiv.org/abs/1802.07108} {\path{arXiv:1802.07108}},
  \href {http://dx.doi.org/10.1103/PhysRevD.97.124065}
  {\path{doi:10.1103/PhysRevD.97.124065}}.

\bibitem{Chakraborty:2015aja}
S.~Chakraborty, K.~Parattu, T.~Padmanabhan, {Gravitational field equations near
  an arbitrary null surface expressed as a thermodynamic identity}, JHEP 10
  (2015) 097.
\newblock \href {http://arxiv.org/abs/1505.05297} {\path{arXiv:1505.05297}},
  \href {http://dx.doi.org/10.1007/JHEP10(2015)097}
  {\path{doi:10.1007/JHEP10(2015)097}}.

\bibitem{Chakraborty:2015hna}
S.~Chakraborty, T.~Padmanabhan, {Thermodynamical interpretation of the
  geometrical variables associated with null surfaces}, Phys. Rev. D 92~(10)
  (2015) 104011.
\newblock \href {http://arxiv.org/abs/1508.04060} {\path{arXiv:1508.04060}},
  \href {http://dx.doi.org/10.1103/PhysRevD.92.104011}
  {\path{doi:10.1103/PhysRevD.92.104011}}.

\bibitem{Damour:1979wya}
T.~Damour, {Quelques proprietes mecaniques, electromagnet iques,
  thermodynamiques et quantiques des trous noir}, Ph.D. thesis, Paris U.,
  VI-VII (1979).

\bibitem{Gourgoulhon:2005ng}
E.~Gourgoulhon, J.~L. Jaramillo, {A 3+1 perspective on null hypersurfaces and
  isolated horizons}, Phys. Rept. 423 (2006) 159--294.
\newblock \href {http://arxiv.org/abs/gr-qc/0503113}
  {\path{arXiv:gr-qc/0503113}}, \href
  {http://dx.doi.org/10.1016/j.physrep.2005.10.005}
  {\path{doi:10.1016/j.physrep.2005.10.005}}.

\bibitem{Padmanabhan:2010rp}
T.~Padmanabhan, {Entropy density of spacetime and the Navier-Stokes fluid
  dynamics of null surfaces}, Phys. Rev. D 83 (2011) 044048.
\newblock \href {http://arxiv.org/abs/1012.0119} {\path{arXiv:1012.0119}},
  \href {http://dx.doi.org/10.1103/PhysRevD.83.044048}
  {\path{doi:10.1103/PhysRevD.83.044048}}.

\bibitem{Kolekar:2011gw}
S.~Kolekar, T.~Padmanabhan, {Action Principle for the Fluid-Gravity
  Correspondence and Emergent Gravity}, Phys. Rev. D 85 (2012) 024004.
\newblock \href {http://arxiv.org/abs/1109.5353} {\path{arXiv:1109.5353}},
  \href {http://dx.doi.org/10.1103/PhysRevD.85.024004}
  {\path{doi:10.1103/PhysRevD.85.024004}}.

\bibitem{Poisson:2009pwt}
E.~Poisson, {A Relativist's Toolkit: The Mathematics of Black-Hole Mechanics},
  Cambridge University Press, Cambridge, U.K., 2009.
\newblock \href {http://dx.doi.org/10.1017/CBO9780511606601}
  {\path{doi:10.1017/CBO9780511606601}}.

\bibitem{Jacobson:1995ab}
T.~Jacobson, {Thermodynamics of space-time: The Einstein equation of state},
  Phys. Rev. Lett. 75 (1995) 1260--1263.
\newblock \href {http://arxiv.org/abs/gr-qc/9504004}
  {\path{arXiv:gr-qc/9504004}}, \href
  {http://dx.doi.org/10.1103/PhysRevLett.75.1260}
  {\path{doi:10.1103/PhysRevLett.75.1260}}.

\bibitem{Eling:2006aw}
C.~Eling, R.~Guedens, T.~Jacobson, {Non-equilibrium thermodynamics of
  spacetime}, Phys. Rev. Lett. 96 (2006) 121301.
\newblock \href {http://arxiv.org/abs/gr-qc/0602001}
  {\path{arXiv:gr-qc/0602001}}, \href
  {http://dx.doi.org/10.1103/PhysRevLett.96.121301}
  {\path{doi:10.1103/PhysRevLett.96.121301}}.

\bibitem{Chirco:2009dc}
G.~Chirco, S.~Liberati, {Non-equilibrium Thermodynamics of Spacetime: The Role
  of Gravitational Dissipation}, Phys. Rev. D 81 (2010) 024016.
\newblock \href {http://arxiv.org/abs/0909.4194} {\path{arXiv:0909.4194}},
  \href {http://dx.doi.org/10.1103/PhysRevD.81.024016}
  {\path{doi:10.1103/PhysRevD.81.024016}}.

\bibitem{Dey:2017fld}
R.~Dey, S.~Liberati, D.~Pranzetti, {Spacetime thermodynamics in the presence of
  torsion}, Phys. Rev. D 96~(12) (2017) 124032.
\newblock \href {http://arxiv.org/abs/1709.04031} {\path{arXiv:1709.04031}},
  \href {http://dx.doi.org/10.1103/PhysRevD.96.124032}
  {\path{doi:10.1103/PhysRevD.96.124032}}.

\bibitem{Baccetti:2013ica}
V.~Baccetti, M.~Visser, {Clausius entropy for arbitrary bifurcate null
  surfaces}, Class. Quant. Grav. 31 (2014) 035009.
\newblock \href {http://arxiv.org/abs/1303.3185} {\path{arXiv:1303.3185}},
  \href {http://dx.doi.org/10.1088/0264-9381/31/3/035009}
  {\path{doi:10.1088/0264-9381/31/3/035009}}.

\bibitem{Parikh:2017aas}
M.~Parikh, A.~Svesko, {Einstein's equations from the stretched future light
  cone}, Phys. Rev. D 98~(2) (2018) 026018.
\newblock \href {http://arxiv.org/abs/1712.08475} {\path{arXiv:1712.08475}},
  \href {http://dx.doi.org/10.1103/PhysRevD.98.026018}
  {\path{doi:10.1103/PhysRevD.98.026018}}.

\bibitem{Jacobson:2015hqa}
T.~Jacobson, {Entanglement Equilibrium and the Einstein Equation}, Phys. Rev.
  Lett. 116~(20) (2016) 201101.
\newblock \href {http://arxiv.org/abs/1505.04753} {\path{arXiv:1505.04753}},
  \href {http://dx.doi.org/10.1103/PhysRevLett.116.201101}
  {\path{doi:10.1103/PhysRevLett.116.201101}}.

\bibitem{Padmanabhan:2002sha}
T.~Padmanabhan, {Classical and quantum thermodynamics of horizons in
  spherically symmetric space-times}, Class. Quant. Grav. 19 (2002) 5387--5408.
\newblock \href {http://arxiv.org/abs/gr-qc/0204019}
  {\path{arXiv:gr-qc/0204019}}, \href
  {http://dx.doi.org/10.1088/0264-9381/19/21/306}
  {\path{doi:10.1088/0264-9381/19/21/306}}.

\bibitem{Kothawala:2007em}
D.~Kothawala, S.~Sarkar, T.~Padmanabhan, {Einstein's equations as a
  thermodynamic identity: The Cases of stationary axisymmetric horizons and
  evolving spherically symmetric horizons}, Phys. Lett. B 652 (2007) 338--342.
\newblock \href {http://arxiv.org/abs/gr-qc/0701002}
  {\path{arXiv:gr-qc/0701002}}, \href
  {http://dx.doi.org/10.1016/j.physletb.2007.07.021}
  {\path{doi:10.1016/j.physletb.2007.07.021}}.

\bibitem{Paranjape:2006ca}
A.~Paranjape, S.~Sarkar, T.~Padmanabhan, {Thermodynamic route to field
  equations in Lancos-Lovelock gravity}, Phys. Rev. D 74 (2006) 104015.
\newblock \href {http://arxiv.org/abs/hep-th/0607240}
  {\path{arXiv:hep-th/0607240}}, \href
  {http://dx.doi.org/10.1103/PhysRevD.74.104015}
  {\path{doi:10.1103/PhysRevD.74.104015}}.

\bibitem{Kothawala:2009kc}
D.~Kothawala, T.~Padmanabhan, {Thermodynamic structure of Lanczos-Lovelock
  field equations from near-horizon symmetries}, Phys. Rev. D 79 (2009) 104020.
\newblock \href {http://arxiv.org/abs/0904.0215} {\path{arXiv:0904.0215}},
  \href {http://dx.doi.org/10.1103/PhysRevD.79.104020}
  {\path{doi:10.1103/PhysRevD.79.104020}}.

\bibitem{Chakraborty:2015wma}
S.~Chakraborty, {Lanczos-Lovelock gravity from a thermodynamic perspective},
  JHEP 08 (2015) 029.
\newblock \href {http://arxiv.org/abs/1505.07272} {\path{arXiv:1505.07272}},
  \href {http://dx.doi.org/10.1007/JHEP08(2015)029}
  {\path{doi:10.1007/JHEP08(2015)029}}.

\bibitem{Friedrich:1998wq}
H.~Friedrich, I.~Racz, R.~M. Wald, {On the rigidity theorem for space-times
  with a stationary event horizon or a compact Cauchy horizon}, Commun. Math.
  Phys. 204 (1999) 691--707.
\newblock \href {http://arxiv.org/abs/gr-qc/9811021}
  {\path{arXiv:gr-qc/9811021}}, \href {http://dx.doi.org/10.1007/s002200050662}
  {\path{doi:10.1007/s002200050662}}.

\bibitem{morales2008second}
E.~M. Morales,
  \href{https://www.theorie.physik.uni-goettingen.de/forschung/qft/theses/dipl/Morfa-Morales.pdf}{On
  a second law of black hole mechanics in a higher derivative theory of
  gravity} (2008).
\newline\urlprefix\url{https://www.theorie.physik.uni-goettingen.de/forschung/qft/theses/dipl/Morfa-Morales.pdf}

\bibitem{Parattu:2015gga}
K.~Parattu, S.~Chakraborty, B.~R. Majhi, T.~Padmanabhan, {A Boundary Term for
  the Gravitational Action with Null Boundaries}, Gen. Rel. Grav. 48~(7) (2016)
  94.
\newblock \href {http://arxiv.org/abs/1501.01053} {\path{arXiv:1501.01053}},
  \href {http://dx.doi.org/10.1007/s10714-016-2093-7}
  {\path{doi:10.1007/s10714-016-2093-7}}.

\bibitem{Racz:2007pv}
I.~Racz, {Stationary Black Holes as Holographs}, Class. Quant. Grav. 24 (2007)
  5541--5572.
\newblock \href {http://arxiv.org/abs/gr-qc/0701104}
  {\path{arXiv:gr-qc/0701104}}, \href
  {http://dx.doi.org/10.1088/0264-9381/24/22/016}
  {\path{doi:10.1088/0264-9381/24/22/016}}.

\bibitem{Smarr:1972kt}
L.~Smarr, {Mass formula for Kerr black holes}, Phys. Rev. Lett. 30 (1973)
  71--73, [Erratum: Phys.Rev.Lett. 30, 521--521 (1973)].
\newblock \href {http://dx.doi.org/10.1103/PhysRevLett.30.71}
  {\path{doi:10.1103/PhysRevLett.30.71}}.

\bibitem{Padmanabhan:2003pk}
T.~Padmanabhan, {Gravitational entropy of static space-times and microscopic
  density of states}, Class. Quant. Grav. 21 (2004) 4485--4494.
\newblock \href {http://arxiv.org/abs/gr-qc/0308070}
  {\path{arXiv:gr-qc/0308070}}, \href
  {http://dx.doi.org/10.1088/0264-9381/21/18/013}
  {\path{doi:10.1088/0264-9381/21/18/013}}.

\bibitem{Padmanabhan:2009kr}
T.~Padmanabhan, {Equipartition of energy in the horizon degrees of freedom and
  the emergence of gravity}, Mod. Phys. Lett. A 25 (2010) 1129--1136.
\newblock \href {http://arxiv.org/abs/0912.3165} {\path{arXiv:0912.3165}},
  \href {http://dx.doi.org/10.1142/S021773231003313X}
  {\path{doi:10.1142/S021773231003313X}}.

\bibitem{Banerjee:2010yd}
R.~Banerjee, B.~R. Majhi, {Statistical Origin of Gravity}, Phys. Rev. D 81
  (2010) 124006.
\newblock \href {http://arxiv.org/abs/1003.2312} {\path{arXiv:1003.2312}},
  \href {http://dx.doi.org/10.1103/PhysRevD.81.124006}
  {\path{doi:10.1103/PhysRevD.81.124006}}.

\bibitem{Banerjee:2010ye}
R.~Banerjee, B.~R. Majhi, S.~K. Modak, S.~Samanta, {Killing Symmetries and
  Smarr Formula for Black Holes in Arbitrary Dimensions}, Phys. Rev. D 82
  (2010) 124002.
\newblock \href {http://arxiv.org/abs/1007.5204} {\path{arXiv:1007.5204}},
  \href {http://dx.doi.org/10.1103/PhysRevD.82.124002}
  {\path{doi:10.1103/PhysRevD.82.124002}}.

\bibitem{Kothawala:2010bf}
D.~Kothawala, {The thermodynamic structure of Einstein tensor}, Phys. Rev. D 83
  (2011) 024026.
\newblock \href {http://arxiv.org/abs/1010.2207} {\path{arXiv:1010.2207}},
  \href {http://dx.doi.org/10.1103/PhysRevD.83.024026}
  {\path{doi:10.1103/PhysRevD.83.024026}}.

\bibitem{padmanabhan2010gravitation}
T.~Padmanabhan, Gravitation: foundations and frontiers, Cambridge University
  Press, New York, United States of America, 2010.

\bibitem{Jacobson:1995uq}
T.~Jacobson, G.~Kang, R.~C. Myers, {Increase of black hole entropy in higher
  curvature gravity}, Phys. Rev. D 52 (1995) 3518--3528.
\newblock \href {http://arxiv.org/abs/gr-qc/9503020}
  {\path{arXiv:gr-qc/9503020}}, \href
  {http://dx.doi.org/10.1103/PhysRevD.52.3518}
  {\path{doi:10.1103/PhysRevD.52.3518}}.

\bibitem{Chatterjee:2011wj}
A.~Chatterjee, S.~Sarkar, {Physical process first law and increase of horizon
  entropy for black holes in Einstein-Gauss-Bonnet gravity}, Phys. Rev. Lett.
  108 (2012) 091301.
\newblock \href {http://arxiv.org/abs/1111.3021} {\path{arXiv:1111.3021}},
  \href {http://dx.doi.org/10.1103/PhysRevLett.108.091301}
  {\path{doi:10.1103/PhysRevLett.108.091301}}.

\bibitem{Majhi:2014hpa}
B.~R. Majhi, {Thermodynamics of Sultana-Dyer Black Hole}, JCAP 05 (2014) 014.
\newblock \href {http://arxiv.org/abs/1403.4058} {\path{arXiv:1403.4058}},
  \href {http://dx.doi.org/10.1088/1475-7516/2014/05/014}
  {\path{doi:10.1088/1475-7516/2014/05/014}}.

\bibitem{Hayward:1993ph}
S.~A. Hayward, {Quasilocal gravitational energy}, Phys. Rev. D 49 (1994)
  831--839.
\newblock \href {http://arxiv.org/abs/gr-qc/9303030}
  {\path{arXiv:gr-qc/9303030}}, \href
  {http://dx.doi.org/10.1103/PhysRevD.49.831}
  {\path{doi:10.1103/PhysRevD.49.831}}.

\bibitem{Prain:2015tda}
A.~Prain, V.~Vitagliano, V.~Faraoni, M.~Lapierre-L\'eonard, {Hawking\textendash
  Hayward quasi-local energy under conformal transformations}, Class. Quant.
  Grav. 33~(14) (2016) 145008.
\newblock \href {http://arxiv.org/abs/1501.02977} {\path{arXiv:1501.02977}},
  \href {http://dx.doi.org/10.1088/0264-9381/33/14/145008}
  {\path{doi:10.1088/0264-9381/33/14/145008}}.

\bibitem{Hayward:1997jp}
S.~A. Hayward, {Unified first law of black hole dynamics and relativistic
  thermodynamics}, Class. Quant. Grav. 15 (1998) 3147--3162.
\newblock \href {http://arxiv.org/abs/gr-qc/9710089}
  {\path{arXiv:gr-qc/9710089}}, \href
  {http://dx.doi.org/10.1088/0264-9381/15/10/017}
  {\path{doi:10.1088/0264-9381/15/10/017}}.

\end{thebibliography}

\end{document}